\documentclass[%
 reprint,
 amsmath,amssymb,
 aps,
]{revtex4-2}

\usepackage{amssymb}
\usepackage{amsmath}
\usepackage{amsfonts}
\usepackage{multirow}
\usepackage{algorithmic}
\usepackage{textcomp}
\usepackage{xcolor}
\usepackage{physics}

\usepackage{graphicx}
\usepackage{dcolumn}
\usepackage{bm}

\newcommand{\eqN}[1]{Eq.~(\ref{#1})} 

\begin{document}

\preprint{APS/123-QED}

\title{Parametric Instability in Discrete Models of Spatiotemporally Modulated Materials}

\author{Jiuda Wu}
\author{Behrooz Yousefzadeh}%
 \email{behrooz.yousefzadeh@concordia.ca}
\affiliation{%
 Department of Mechanical, Industrial and Aerospace Engineering, Concordia University\\
 1455 De Maisonneuve Blvd. W., Montreal, Quebec, H3G 1M8, Canada
}%




\date{\today}

\begin{abstract}
We investigate the phenomenon of parametric instability in discrete models of spatiotemporally modulated materials. These materials are celebrated in part because they exhibit nonreciprocal transmission characteristics. However, parametric instability may occur for strong modulations, or occasionally even at very small modulation amplitudes, and prevent the safe operation of spatiotemporally modulated devices due to an exponential growth in the response amplitude. We use Floquet theory to conduct a detailed computational investigation of parametric instability. We explore the roles of modulation parameters (frequency, amplitude, wavenumber), the number of modulated units, and damping on the stability of the system. We highlight the pivotal role of spatial modulation in parametric instability, a feature that is predominantly overlooked in this context. We use the perturbation method to obtain analytical expressions for modulation frequencies at which the response becomes unstable. We hope that our findings enable and inspire new applications of spatiotemporally modulated materials that operate at higher amplitudes. 
\end{abstract}

\maketitle


\section{Introduction} 
\label{sec_intro}

Spatiotemporal modulation of the effective material properties of a system is one established way to realize nonreciprocal transmission of mechanical or acoustic waves~\cite{KLurie_2017,HNassar_2020}. In this context, spatiotemporal modulation refers to periodic changes (often harmonic) in the material properties of a waveguide, material or device in both space and time. The nonreciprocal transmission characteristics of spatiotemporally modulated materials have been a key factor in the great attention they have received in recent years. 

The underlying mechanisms that lead to nonreciprocal propagation in spatiotemporally modulated materials are relatively well understood by now~\cite{HNassar_2018,deymier_2019,NJP2016,PRA2022_Nouh}. A widely featured dynamic characteristic of these systems is the appearance of direction-dependent frequency gaps in the dispersion diagram, which leads to unidirectional propagation of waves through the system. In finite systems, this leads to a large difference between the energies transmitted in opposite directions: a large energy bias. 

In systems with very few spatiotemporally modulated units, the energy bias is often very small and nonreciprocity manifests primarily as a difference in the transmitted phases instead~\cite{paper1}. To increase the energy bias in short systems, one can increase the number of modulated units or the modulation amplitude. The influence of adding more modulated units on the transmission characteristics can be investigated using the theoretical frameworks that already exist in the literature. This is no longer the case for strongly modulated systems. 

Increasing the strength of modulations can change the spectral contents of the response, specifically by making the contributions from sidebands more significant and by shifting the resonance frequencies; a detailed analysis of these effects is available elsewhere~\cite{paper1}. More importantly, strong modulations can result in parametric instability, which leads to unbounded growth of the response amplitude in time~\cite{Champneys,IKovacic_2018}. This is a critical feature of strongly modulated systems because it can compromise safe operation of modulated devices or cause device failure. 

The objective of this work is to investigate the phenomenon of parametric instability in spatiotemporally modulated systems. We focus exclusively on discrete models of spatiotemporally modulated materials. To a great extent, this choice is motivated by the models associated with experimental realization of spatiotemporal modulations. Spatiotemporal modulation, a type of parametric excitation, is often achieved at discrete points throughout the structure, for example by piezoelectric patches~\cite{GTrainiti_2019,RThomes_2023} or magnetic forces~\cite{YWang_2018,SWan_2022}. Discrete models are therefore developed for their analysis, especially in the case of finite systems. 

Parametric excitation occurs in numerous mechanical systems when a displacement-dependent forcing is present, perhaps most famously in a pendulum with a moving base~\cite[Ch. 5]{ANayfeh_1979} or used to explain how to get a swing in motion~\cite{swing}. The study of parametrically excited systems dates back to the nineteenth century~\cite{MathieuEq_hist}, with the works of Mathieu~\cite{Mathieu_1868} and Rayleigh~\cite{JRayleigh_1883} among the early contributions in mechanical vibrations. In the present century, parametric excitation and the associated amplification effect are widely utilized in the operation of MEMS sensors and actuators~\cite{JRhoads_2008,SShaw_2018}. 

Parametric instabilities occur when the system can no longer maintain a bounded response amplitude above a threshold of modulation amplitude -- this threshold can be infinitesimally small under certain conditions. The ensuing exponential growth of the response amplitude is detrimental to a system's ability to carry out its intended operation and often leads to failure. Understanding of parametric stability is therefore crucial in applications ranging from aerospace rotors~\cite{aero1, aero2} and wind turbines~\cite{BFeeny_2020} to machining chatter~\cite{chatter1, chatter2} and control systems~\cite{control1}. 

Spatiotemporally modulated materials can be modeled as coupled oscillators subject to parametric excitation. The corresponding mathematical model is a system of coupled Mathieu equations, with a phase term that corresponds to spatial modulation. Despite the vast literature on the Mathieu equation, there are relatively fewer studies on coupled Mathieu equations, and we have found that parametric stability is rarely discussed in the context of spatiotemporally modulated systems~\cite{nassar_python,YJin_2024}. In particular, the influence of spatial modulations (modulation wavenumber) on parametric stability remains to be investigated. Our goal in this work is to contribute to filling this gap in the literature. 

We refer to the modulation frequencies that can lead to unstable response for infinitesimally small modulation amplitudes as the unstable modulation frequencies (UMFs) -- this only occurs in the absence of energy dissipation. One of the early studies on parametric stability in a system of $n$ undamped Mathieu equations found that the response remains stable for sufficiently small modulation amplitudes unless the modulation frequency is equal to a combination of any two natural frequencies of the system in the absence of modulations~\cite[Ch. 4]{LCesari_1963}. Specifically, $|\Omega_{n,j_1}\pm\Omega_{n,j_2}|/\beta>0$ are identified as potential UMFs, where $j_{1,2}\in\{1,2,\cdots,n\}$, $\beta \in \mathbb{N}$, and $\Omega_{n,j_{1,2}}$ denotes an unmodulated natural frequency. An independent study based on  perturbation analysis reported the same result~\cite{CHsu_1963}. 

It has long been established to use stability diagrams (stability charts) to graphically present the dependence of parametric stability on modulation amplitude and frequency~\cite{Ince_1927}. Despite the extensive research on parametric stability of Mathieu's equation and its various extended forms, systems with more degrees of freedom (DoF) have received much less attention in comparison. For example, the UMFs have been studied in coupled modulated systems (2-DoF)~\cite{Unstb_freq2,Unstb_freq3}, for which the presence of a phase shift between the two modulations influences the stability but not the  UMFs~\cite{Unstb_freq2}. When the parametric excitation is applied to individual degrees of freedom independently (not appearing in coupling), only $\Omega_{2,j_1}+\Omega_{2,j_2}$ are identified as UMFs~\cite{Unstb_freq1}. In a study on a 3-DoF system, the stability diagram shows an increase in the  number of unstable regions, with the UMFs located at $(\Omega_{3,j_1}+\Omega_{3,j_2})/\beta$~\cite{Unstb_freq3}. Nevertheless, we could not find a systematic study of parametric instability that can be readily applied in the context of spatiotemporally modulated systems. In particular, the influence of spatial modulations and the number of modulated units on parametric stability remains unexplored for the most part. 

In this work, we present a detailed computational analysis of parametric instability in spatiotemporally modulated systems. We use Floquet theory to determine the stability of response of the system. We also perform a perturbation analysis of the UMFs that incorporates the influence of modulation phase (wavenumber). The stability of long modulated systems is investigated and discussed in more detail than we could find in the literature. For both short and longer systems, we highlight contiguous regions of stable response in the stability diagrams that cover a significant range of modulation amplitudes and appear at lower values of modulation frequency in stability diagrams. These wide ranges of modulation amplitude support the design of slowly modulated systems with high modulation amplitudes. 

When a vibrating system is subject to simultaneous external and parametric excitation with the modulation frequency locked at twice the frequency of the external drive, we can observe unbounded amplitude in the steady-state response of the system even in the presence of damping. This phenomenon is called parametric amplification~\cite{JRhoads_2008}, and is distinct from the instability mechanism we explore in this work. We also note that wavenumber bandgaps may appear in dispersion curves, indicating the appearance of standing waves with exponentially growing amplitudes, especially when the modulation frequency is high~\cite{GTrainiti_2019,BKim_2023,CChong_2024}. 

We present the discrete model of a one-dimensional spatiotemporally modulated system in Section~\ref{sec_formu}. Section~\ref{sec_approach} introduces the approaches used to determine the stability of the response: Floquet theory and perturbation analysis. Stability diagrams for short systems are presented in Section~\ref{sec_short}. In Section~\ref{sec_phi}, we investigate the influence of spatial modulation on parametric stability. The stability of long modulated systems is explored in Section~\ref{sec_long}. The influence of damping on stability is analyzed in Section~\ref{sec_zeta}. Section~\ref{sec_conclusions} summarizes our findings. 

\section{Problem formulation} \label{sec_formu}
\begin{figure}[htb]
\centerline{\includegraphics[scale=0.45]{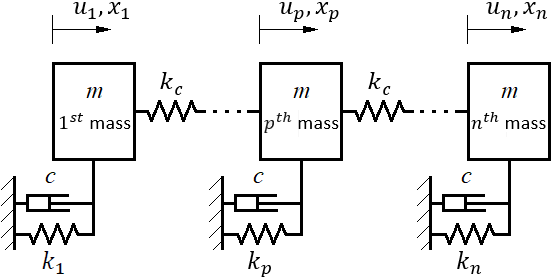}}
\caption{Schematic representation of the modulated system with $n$
DoF.}\label{fig_nDoF}
\end{figure}
We consider a discrete model of one-dimensional spatiotemporally modulated materials in this work. Fig.~\ref{fig_nDoF} shows the schematic of this model. The system is composed of identical masses, linear coupling springs, viscous dampers, and modulated grounding springs. For each mass, only the longitudinal rectilinear motion is considered as a degree of freedom (DoF). The stiffness coefficient of the grounding spring in the $p~\!$-th modulated unit is expressed as $k_p(\tau)=k_{g,DC}+k_{g,AC}\cos{\left( \omega_mt-\phi_p \right)}$, where $\phi_p=\left(p-1\right)\phi$ for $p=1,2,\cdots,n$. Parameters $\omega_m$ and $k_{g,AC}$ are the modulation frequency and amplitude, respectively. Parameter $\phi$ represents the spatial modulation along the system. This is the same as the modulation wavenumber. We refer to $\phi$ as the modulation phase in this work because of our emphasis on short systems. 

The equations of motion for the modulated system in Fig.~\ref{fig_nDoF} are first nondimensionalized; see Appendix~\ref{appendix:nondimensionalization}. The nondimensional equations of motion for the $p~\!$-th mass of the system are:
\begin{equation}
\label{eq_EoM_p}
\ddot{x}_p +2 \zeta \dot{x}_p + x_p \left[1 + K_m \cos{\left( \Omega_m \tau - \phi_p \right)} \right] + K_c \Delta^2_p = 0, 
\end{equation}
where the overdot represents differentiation with respect to nondimentional time $\tau$. The difference terms representing coupling are $\Delta^2_1 = x_1 - x_2$ and $\Delta^2_n = x_n - x_{n-1}$ at two ends of the system; $\Delta^2_p = 2 x_p - x_{p+1} - x_{p-1}$ elsewhere. We will continue with the nondimensional equations. 
 
\begin{figure}[htb]
\centerline{\includegraphics[scale=0.4]{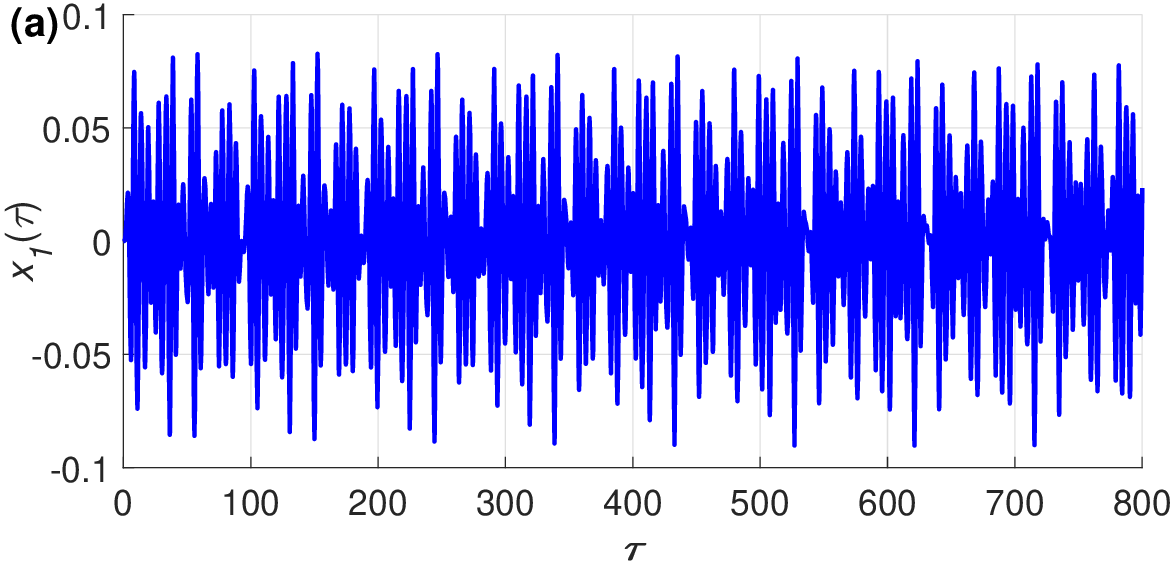}}
\centerline{\includegraphics[scale=0.4]{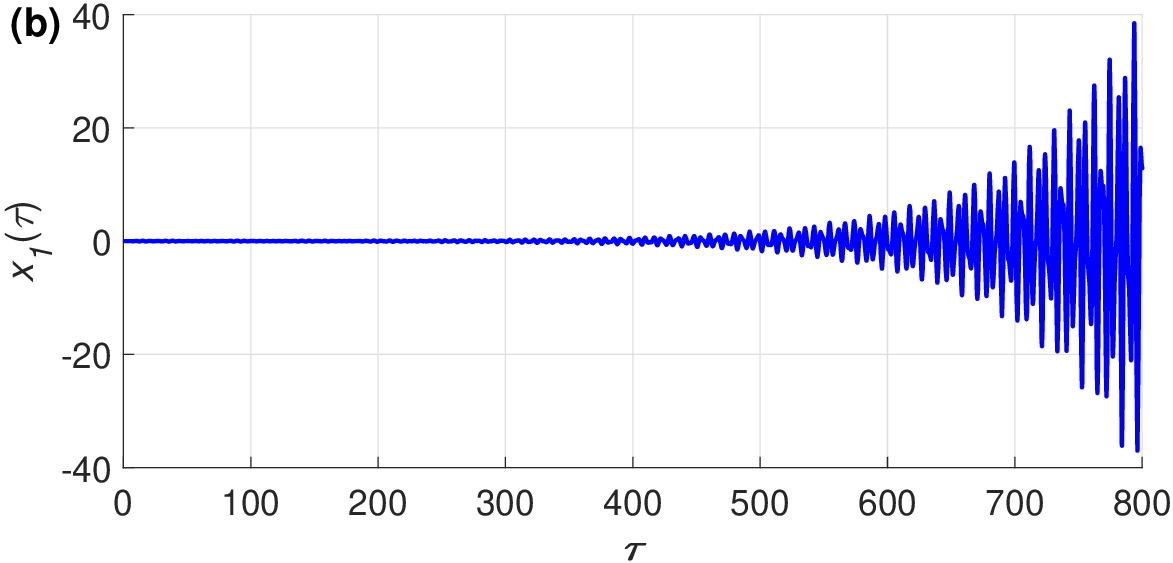}}
\caption{Displacements of the first masses in two modulated systems with $K_c=0.6$, $\phi=0.5\pi$ and $\zeta=0$. (a) $n=3$, $\Omega_m=2.6$ and $K_m=0.2$; this scenario falls inside a stable region in Fig.~\ref{fig_STB_1}(b). (b) $n=5$, $\Omega_m=2.6$ and $K_m=0.1$; this scenario falls inside an unstable region in Fig.~\ref{fig_STB_1}(b). The initial conditions for both examples are: $\dot{x}_n\left(0\right)=0.1$, $x_n\left(0\right)=x_p\left(0\right)=\dot{x}_p\left(0\right)=0$ for $1\leq p \leq n-1$.}
\label{fig_time_1}
\end{figure}

Fig.~\ref{fig_time_1} shows the time-domain response of Eq.(\ref{eq_EoM_p}) for two different sets of system parameters. Panel (a) shows a typical stable response, characterized by its quasiperiodic nature and constant amplitude (energy). Panel (b) shows a typical unstable response, which is characterized by an exponentially growing amplitude. 

\section{Approaches to determine parametric stability} \label{sec_approach}

\subsection{Floquet theory: Direct computation}
\label{subsec1}

Eq.(\ref{eq_EoM_p}) can be recast as a system of linear ordinary differential equations: 
\begin{equation}
\label{eq_1}
\dv{}{\tau}\,\underline{\mathit{X}}\left(\tau\right)=\underline{\underline{\mathsf{A}}}\left(\tau\right)\,\,\underline{\mathit{X}}\left(\tau\right) ,
\end{equation}
where $\underline{\mathit{X}}=\{\dot{\underline{x}},\underline{x}\}^T$ and $\underline{x}=\{x_1,x_2,\cdots,x_n\}^T$. The matrix of coefficients is periodic in time, $\underline{\underline{\mathsf{A}}}\left(\tau\right) = \underline{\underline{\mathsf{A}}}\left(\tau+T_E\right)$, with $T_E=2\pi/\Omega_m$ in this case. Vector and matrix variables are indicated with single and double underlines. Floquet theory describes the conditions for stability (boundedness) of the solutions, $X(\tau)$, based on the principal matrix of the system~\cite{GFloquet_1883}. 

The principal matrix of Eq.(\ref{eq_EoM_p}), denoted by $\underline{\underline{\Psi}}$, satisfies $\underline{\underline{\mathit{X}}}\left(T_E\right) = \underline{\underline{\Psi}} \, \underline{\underline{\mathit{X}}}\left(0\right)$, where $\underline{\underline{\mathit{X}}}\left(0\right)$ contains an independent set of initial conditions (the identity matrix is the most common choice). 
The eigenvalues of the principal matrix are crucial to determine the stability of the response of the system. Because there is no explicit analytical solution for~\eqN{eq_1}, or for Mathieu's equation, the stability of the solutions are typically computed numerically. Several approximation methods have also been developed to obtain the principal matrix~\cite{JRayleigh_1887,ANayfeh_1979,SSinha_1991}. In addition to approaches that focus on the principal matrix, the stability of Mathieu's equation can be determined by the harmonic-balance method and the method of multiple scales~\cite{TInsperger_2002,IKovacic_2018}. 

The $2n\times2n$  matrix of coefficients in ~\eqN{eq_1} is: 
\begin{equation}
\underline{\underline{\mathsf{A}}}(\tau)=\begin{bmatrix}
\underline{\underline{\mathsf{D}}} & \underline{\underline{\mathsf{C}}}(\tau)\\
\underline{\underline{I}} & \underline{\underline{O}}
\end{bmatrix}
\end{equation}
where $\underline{\underline{\mathsf{D}}}$, $\underline{\underline{\mathsf{C}}}(\tau)$, $\underline{\underline{I}}$ and $\underline{\underline{O}}$ are all $n\times n$ matrices. $\underline{\underline{O}}$ is a zero matrix, $\underline{\underline{I}}$ is an identity matrix and $\underline{\underline{\mathsf{D}}}=-2\zeta\underline{\underline{I}}$. Matrix $\underline{\underline{\mathsf{C}}}(\tau)$ can be written as:
\begin{equation}
\underline{\underline{\mathsf{C}}}(\tau)=\begin{bmatrix}
\mathsf{B}_1(\tau) & K_c & 0 & \cdots & 0\\
K_c & \mathsf{B}_2(\tau) & K_c & \cdots & 0\\
0 & K_c & \mathsf{B}_3(\tau) & \cdots & 0\\
\vdots & \vdots & \vdots & \ddots  & \vdots\\
0 & 0 & 0 & \cdots & \mathsf{B}_n(\tau)
\end{bmatrix}
\end{equation}
where the diagonal terms $\mathsf{B}_p(\tau)=-2K_c-[1+K_m\cos(\Omega_m\tau-\phi_p)]$ for any value of $p$, except at the two ends where $\mathsf{B}_1(\tau)=-K_c-[1+K_m\cos(\Omega_m\tau)]$ and $\mathsf{B}_n(\tau)=-K_c-[1+K_m\cos(\Omega_m\tau-\phi_n)]$. The elements in the first super diagonal and the first subdiagonal of $\underline{\underline{\mathsf{C}}}(\tau)$ are equal to $K_c$. The remaining elements of $\underline{\underline{\mathsf{C}}}(\tau)$ are zero. 

We define $2n$ vectors of initial conditions that can form an identity matrix: 
\begin{equation}
\begin{bmatrix}
\underline{\mathit{X}}_1(0) & \underline{\mathit{X}}_2(0) & \cdots & \underline{\mathit{X}}_{2n}(0)
\end{bmatrix} = \underline{\underline{I}}.
\end{equation}
The response of the modulated system at $\tau=T_E$, computed for each set of initial conditions, forms a matrix called the principal matrix: 
\begin{equation}
\underline{\underline{\mathsf{E}}}=\begin{bmatrix}
\underline{\mathit{X}}_1(T_E) & \underline{\mathit{X}}_2(T_E) & \cdots & \underline{\mathit{X}}_{2n}(T_E)
\end{bmatrix}. 
\end{equation}
Floquet theory states that the response of the system becomes unstable (unbounded) if any eigenvalue of $\underline{\underline{\mathsf{E}}}$ has its modulus larger than unity. 
In this work, we compute the principal matrices using the fourth-order Runge-Kutta method with Gill coefficients~\cite{SGill_1951,BCarnahan_1969}. All the stability diagrams reported in subsequent sections are computed in this way. 

\subsection{Perturbation method: Predicting unstable modulation frequencies}
\label{sec:perturbation}

Our goal in this section is to obtain analytical expressions for modulation frequencies that lead to unbounded response in the presence of (infinitesimally) small modulation amplitude. We refer to these frequencies as the unstable modulation frequencies, UMFs. We decompose the steady-state response of the system into its constituent modes to facilitate the analysis based on the method of multiple scales. 

In the absence of energy loss ($\zeta=0$), the response of the modulated system governed by~\eqN{eq_EoM_p} can be expressed as a Fourier series:
\begin{equation}
\label{eq_xp_solusion}
x_p\left(\tau\right) = \sum^{n}_{q=1} \sum^{\infty}_{\kappa=-\infty} X_{p;q;\kappa} \left(\tau\right) e^{i\left( \Omega_{n,q} + \kappa \Omega_m \right) \tau}, 
\end{equation}
where $\Omega_{n,q}$ is the $q~\!$-th natural frequency of the unmodulated system. 
An alternative formulation based on the harmonic balance method may be used for systems with one degree of freedom~\cite{TInsperger_2002,TInsperger_2003}. We found that this formulation results in inaccurate prediction of the stability charts for systems with multiple degrees of freedom; see Appendix~\ref{appendix:harmonic_balancing}. 

To keep track of the parametric instabilities that occur at lower modulation frequencies more easily, we introduce the parameter $\beta$ and rewrite $x_p\left(\tau\right)$ as:
\begin{align}
\label{eq_xp_y}
x_p\left(\tau\right) &= \sum^{\infty}_{\beta=1} w_{p;\beta} \left(\tau\right), \\
w_{p;\beta}\left(\tau\right) &= \sum^{n}_{q=1} \sum^{\infty}_{\kappa=-\infty} W_{p;\beta;q;\kappa} \left(\tau\right) e^{i\left( \Omega_{n,q} + \kappa \beta \Omega_m \right) \tau},
\end{align}
where $\beta \in \mathbb{N}$. If any $w_{p;\beta}\left(\tau\right)$ is unbounded, then $x_p\left(\tau\right)$ becomes unbounded too. 
Therefore, we check the boundedness of $w_{p;\beta}\left(\tau\right)$ to determine the parametric stability of $x_p\left(\tau\right)$. $w_{p;\beta}\left(\tau\right)$ satisfies the equation:
\begin{equation}
\label{eq_EoM_yp}
\ddot{w}_{p;\beta} + w_{p;\beta} \left[1 + \epsilon \cos{\left( \beta \Omega_m \tau - \phi_p \right)} \right] + K_c \delta^2_p = 0, 
\end{equation}
where $\delta^2_1 = w_{1;\beta} - w_{2;\beta}$ and $\delta^2_n = w_{n;\beta} - w_{n-1;\beta}$ at the two ends, and $\delta^2_p = 2 w_{p;\beta} - w_{p-1;\beta} - w_{p+1;\beta}$ elsewhere. Parameter $\epsilon = K_m$ is used here to indicate a small value of $K_m$. We note that the unmodulated natural frequencies of~\eqN{eq_EoM_yp} are the same as those of~\eqN{eq_EoM_p}. 

We first decouple the unmodulated terms in~\eqN{eq_EoM_yp}. The mode shape for $\Omega_{n,q}$ is denoted by the vector $\underline{\mathrm{W}}_{q;\beta}=\left\{\mathrm{w}_{q;\beta;1},\mathrm{w}_{q;\beta;2},\cdots,\mathrm{w}_{q;\beta;n}\right\}^T$, where $q=1,2,\cdots,n$. 
In the decoupled (modal) space, the displacement is expressed as $z_{p;{\beta}}=\underline{\mathrm{W}}_{p;\beta}^T \underline{w}_{\beta}$, where $\underline{w}_{\beta}=\left\{w_{1;\beta},w_{2;\beta},\cdots,w_{n;\beta}\right\}^T$. 
The $p$-th decoupled (modal) equation of motion to replace \eqN{eq_EoM_yp} is:
\begin{equation}
\label{eq_EoM_decp_p}
\ddot{z}_{p;\beta} + \Omega_{n,p}^2 z_{p;\beta} + \epsilon \sum^{n}_{q=1} \mathrm{w}_{p;\beta;q} w_{q;\beta} \cos{\left(\beta \Omega_m \tau - \phi_{q}\right)} = 0. 
\end{equation}
Here, $w_{q;\beta}$ can be written out using the inverse transformation $\underline{w}_{\beta}=\underline{\underline{\mathrm{W}}}_{\beta}^{-1} \underline{Z}_{\beta}$, where $\underline{Z}_{\beta} = \left\{z_{1;\beta},z_{2;\beta},\cdots,z_{n;\beta}\right\}^T$ and $\underline{\underline{\mathrm{W}}}_{\beta} = \left[\underline{\mathrm{W}}_{1;\beta}^T;\underline{\mathrm{W}}_{2;\beta}^T;\cdots;\underline{\mathrm{W}}_{n;\beta}^T\right]$. 

The stability of the response of the system is determined by whether the amplitude of $z_{p;\beta}$ becomes unbounded. We present this analysis in detail for the 2-DoF system. Hereafter, we will drop the subscript $\beta$ in $z_{p;\beta}$ for simplicity. 

\subsubsection{2-DoF modulated systems}
\label{sec:2DoF}

The decoupled equations of motion for the 2-DoF system are:
\begin{subequations}
\label{eq_EoM_2DoF}
\begin{align}
\label{eq_EoM_2DoF_1}
\ddot{z}_1 &+ \Omega_{2,1}^2 z_1 + \frac{\epsilon}{2} \left( z_1 + z_2\right) \cos{\beta \Omega_m \tau } \nonumber \\
&+ \frac{\epsilon}{2} \left( z_1 - z_2\right) \cos{\left(\beta \Omega_m \tau - \phi\right)} = 0, \vspace{-1.8mm}
\end{align}
\begin{align}
\label{eq_EoM_2DoF_2}
\ddot{z}_2 &+ \Omega_{2,2}^2 z_2 + \frac{\epsilon}{2} \left( z_1 + z_2\right) \cos{\beta \Omega_m \tau } \nonumber \\
&- \frac{\epsilon}{2} \left( z_1 - z_2\right) \cos{\left(\beta \Omega_m \tau - \phi\right)} = 0,
\end{align}
\end{subequations}
where $\Omega_{2,1}=1$ and $\Omega_{2,2}=\sqrt{1+2K_c}$. To employ the method of multiple scales, we define two (slow and fast) time variables, $\mu=\Omega_m \tau$ and $\nu = \epsilon \Omega_m \tau$. \eqN{eq_EoM_2DoF} is therefore rewritten in the new time variables as:
\begin{subequations}
\label{eq_EoM_2DoF_pert}
\begin{align}
\label{eq_EoM_2DoF_pert_1}
\Omega_m^2 &\pdv[2]{z_1}{\mu} + 2\epsilon\Omega_m^2 \pdv{z_1}{\mu}{\nu} + \epsilon^2 \Omega_m^2 \pdv[2]{z_1}{\nu} + \Omega_{2,1}^2 z_1 \nonumber\\
&+ \frac{\epsilon}{2} \left( z_1 + z_2\right) \cos{\beta \mu } + \frac{\epsilon}{2} \left( z_1 - z_2\right) \cos{\left(\beta \mu - \phi\right)} = 0, 
\end{align}
\vspace{-6mm}
\begin{align}
\label{eq_EoM_2DoF_pert_2}
\Omega_m^2 &\pdv[2]{z_2}{\mu} + 2\epsilon\Omega_m^2 \pdv{z_2}{\mu}{\nu} + \epsilon^2 \Omega_m^2 \pdv[2]{z_2}{\nu} + \Omega_{2,2}^2 z_2 \nonumber\\
&+ \frac{\epsilon}{2} \left( z_1 + z_2\right) \cos{\beta \mu } - \frac{\epsilon}{2} \left( z_1 - z_2\right) \cos{\left(\beta \mu - \phi\right)} = 0.
\end{align}
\end{subequations}
$z_p$ is expanded into a power series of $\epsilon$:
\begin{equation}
\label{eq_zp_series}
z_p\left( \mu,\nu \right) = z_{p,0}\left( \mu,\nu \right) + \epsilon z_{p,1}\left( \mu,\nu \right) + O\left( \epsilon^2 \right).
\end{equation}
We substitute~\eqN{eq_zp_series} into~\eqN{eq_EoM_2DoF_pert} and neglect the terms of $O\left( \epsilon^2 \right)$. Equating the coefficients of $\epsilon^0$ and $\epsilon^1$ gives two sets of equations:
\begin{equation}
\label{eq_EoM_2DoF_pert_e0}
\pdv[2]{z_{j,0}}{\mu} + \left(\frac{\Omega_{2,j}}{\Omega_m}\right)^2 z_{j,0}  = 0,
\end{equation}
where $j \in \{1,2\}$, and 
\begin{widetext}
\begin{subequations}
\label{eq_EoM_2DoF_pert_e1}
\begin{align}
\pdv[2]{z_{1,1}}{\mu} + \left(\frac{\Omega_{2,1}}{\Omega_m}\right)^2 z_{1,1}  = -2 \pdv{z_{1,0}}{\mu}{\nu} - \frac{1}{2\Omega_m^2} \left( z_{1,0} + z_{2,0}\right) \cos{\beta \mu } - \frac{1}{2\Omega_m^2} \left( z_{1,0} - z_{2,0}\right) \cos{\left(\beta \mu - \phi\right)}, 
\end{align}
\vspace{-6mm}
\begin{align}
\pdv[2]{z_{2,1}}{\mu} + \left(\frac{\Omega_{2,2}}{\Omega_m}\right)^2 z_{2,1}  = -2 \pdv{z_{2,0}}{\mu}{\nu} - \frac{1}{2\Omega_m^2} \left( z_{1,0} + z_{2,0}\right) \cos{\beta \mu } + \frac{1}{2\Omega_m^2} \left( z_{1,0} - z_{2,0}\right) \cos{\left(\beta \mu - \phi\right)}.
\end{align}
\end{subequations}
\end{widetext}
The general solution for~\eqN{eq_EoM_2DoF_pert_e0} is:
\begin{equation}
\label{eq_EoM_2DoF_e0_solu}
z_{j,0}\left( \mu,\nu \right) = A_j\left( \nu \right) \cos{\frac{\Omega_{2,j}}{\Omega_m} \mu} + B_j\left( \nu \right) \sin{\frac{\Omega_{2,j}}{\Omega_m} \mu}.
\end{equation}
Substituting~\eqN{eq_EoM_2DoF_e0_solu} into~\eqN{eq_EoM_2DoF_pert_e1} gives: 
\begin{widetext}
\begin{equation}
\label{eq_EoM_2DoF_e1_freqs}
\pdv[2]{z_{j,1}}{\mu} + \left(\frac{\Omega_{2,j}}{\Omega_m}\right)^2 z_{j,1}  = \mathcal{T}_j \left[\frac{\Omega_{2,j}}{\Omega_m}\right] + \mathcal{T}_j \left[\beta+\frac{\Omega_{2,1}}{\Omega_m}\right] + \mathcal{T}_j \left[\beta-\frac{\Omega_{2,1}}{\Omega_m}\right] + \mathcal{T}_j \left[\beta+\frac{\Omega_{2,2}}{\Omega_m}\right] + \mathcal{T}_j \left[\beta-\frac{\Omega_{2,2}}{\Omega_m}\right], 
\end{equation} 
\end{widetext}
where $\mathcal{T}_j\left[\omega\right]$ represents the sum of all harmonic terms at frequency $\omega$ in the $j$-th equation. See Appendix~\ref{appendix:harmonic_terms} for the expressions of each $\mathcal{T}_j\left[\omega\right]$. 

The forcing terms in~\eqN{eq_EoM_2DoF_e1_freqs} can result in unbounded growth of $z_{i,1}$, which leads to unstable response of the system. Therefore, stability is determined by considering the resonance condition of these terms. In perturbation theory, this process is known as removing the {\it secular} terms. 

If neither $|\beta\pm\Omega_{2,1}/\Omega_m|$ nor $|\beta\pm\Omega_{2,2}/\Omega_m|$ is equal to $\Omega_{2,j}/\Omega_m$ for both $j=1$ and $j=2$, then $\mathcal{T}_j\left[\Omega_{2,j}/\Omega_m\right]$ is the only secular term in~\eqN{eq_EoM_2DoF_e1_freqs}. Removing this term results in:
\begin{equation}
\dv{A_j}{\nu} = \dv{B_j}{\nu} = 0.
\end{equation}
Both $z_{j,0}$ and $z_{j,1}$ are bounded when $A_j(\nu)$ and $B_j(\nu)$ are constant. Thus, parametric instability may occur when one of $|\beta\pm\Omega_{2,j_1}/\Omega_m|$ is equal to $\Omega_{2,j_2}/\Omega_m$, where $j_{1,2}\in\{1,2\}$. 

When $\Omega_m=2\Omega_{2,j}/\beta$, both $\mathcal{T}_j\left[\Omega_{2,j}/\Omega_m\right]$ and $\mathcal{T}_j\left[\beta-\Omega_{2,j}/\Omega_m\right]$ include secular terms in the $j$-th equation of~\eqN{eq_EoM_2DoF_e1_freqs}. The removal of these two sets of resonance terms gives:
\begin{equation}
\dv{A_j}{\nu} = -\frac{1+\cos{\phi}}{8 \Omega_{2,j} \Omega_m} B_j + \frac{\sin{\phi}}{8 \Omega_{2,j} \Omega_m} A_j,\nonumber
\end{equation}\vspace{-3mm}
\begin{equation} 
\dv{B_j}{\nu} = -\frac{1+\cos{\phi}}{8 \Omega_{2,j} \Omega_m} A_j - \frac{\sin{\phi}}{8 \Omega_{2,j} \Omega_m} B_j,\nonumber
\end{equation}
and 
\begin{equation}
\dv[2]{A_j}{\nu} = \frac{1+\cos{\phi}}{32 \Omega_{2,j}^2 \Omega_m^2} A_j,\quad \dv[2]{B_j}{\nu} = \frac{1+\cos{\phi}}{32 \Omega_{2,j}^2 \Omega_m^2} B_j.
\end{equation}
When $\cos{\phi}>-1$, both $A_j(\nu)$ and $B_j(\nu)$ grow exponentially, making $\Omega_m=2\Omega_{2,j}/\beta$ a UMF. Otherwise, both $A_j(\nu)$ and $B_j(\nu)$ are constant when $\cos{\phi}=-1$ and the response remains stable.

When $\Omega_m=(\Omega_{2,1}+\Omega_{2,2})/\beta$, we have $\Omega_{2,1}/\Omega_m=\beta-\Omega_{2,2}/\Omega_m$ and $\Omega_{2,2}/\Omega_m=\beta-\Omega_{2,1}/\Omega_m$. Thus, both $\mathcal{T}_1\left[\Omega_{2,1}/\Omega_m\right]$ and $\mathcal{T}_1\left[\beta-\Omega_{2,2}/\Omega_m\right]$ include resonant terms in~\eqN{eq_EoM_2DoF_e1_freqs} with $j=1$. In addition, both $\mathcal{T}_2\left[\Omega_{2,2}/\Omega_m\right]$ and $\mathcal{T}_2\left[\beta-\Omega_{2,1}/\Omega_m\right]$ include resonance terms in~\eqN{eq_EoM_2DoF_e1_freqs} with $j=2$. The removal of these four sets of resonance terms gives:
\begin{equation}
\dv{A_1}{\nu} = -\frac{1-\cos{\phi}}{8 \Omega_{2,1} \Omega_m} B_2 - \frac{\sin{\phi}}{8 \Omega_{2,1} \Omega_m} A_2, \nonumber
\end{equation}\vspace{-3mm}
\begin{equation}
\dv{B_1}{\nu} = -\frac{1-\cos{\phi}}{8 \Omega_{2,1} \Omega_m} A_2 + \frac{\sin{\phi}}{8 \Omega_{2,1} \Omega_m} B_2,\nonumber
\end{equation}\vspace{-3mm}
\begin{equation}
\dv{A_2}{\nu} = \frac{-1+\cos{\phi}}{8 \Omega_{2,2} \Omega_m} B_1 - \frac{\sin{\phi}}{8 \Omega_{2,2} \Omega_m} A_1,\nonumber
\end{equation}\vspace{-3mm}
\begin{equation} 
\dv{B_2}{\nu} = \frac{-1+\cos{\phi}}{8 \Omega_{2,2} \Omega_m} A_1 + \frac{\sin{\phi}}{8 \Omega_{2,2} \Omega_m} B_1,\nonumber
\end{equation}
and 
\begin{equation}
\dv[2]{A_j}{\nu} = \frac{1-\cos{\phi}}{32 \Omega_{2,1} \Omega_{2,2} \Omega_m^2} A_j, \dv[2]{B_j}{\nu} = \frac{1-\cos{\phi}}{32 \Omega_{2,1} \Omega_{2,2} \Omega_m^2} B_j.
\end{equation}
When $-1\le\cos{\phi}<1$, both $A_j(\nu)$ and $B_j(\nu)$ grow exponentially, making $\Omega_m=(\Omega_{2,1}+\Omega_{2,2})/\beta$ a UMF. Otherwise, both $A_j(\nu)$ and $B_j(\nu)$ are constant when $\cos{\phi}=1$ and the response remains stable. 

When $\Omega_m=(\Omega_{2,2}-\Omega_{2,1})/\beta$, we have $\Omega_{2,1}/\Omega_m=\Omega_{2,2}/\Omega_m-\beta$ and $\Omega_{2,2}/\Omega_m=\beta+\Omega_{2,1}/\Omega_m$. Thus, both $\mathcal{T}_1\left[\Omega_{2,1}/\Omega_m\right]$ and $\mathcal{T}_1\left[\beta-\Omega_{2,2}/\Omega_m\right]$ include resonance terms in~\eqN{eq_EoM_2DoF_e1_freqs} with $j=1$. Meanwhile, both $\mathcal{T}_2\left[\Omega_{2,2}/\Omega_m\right]$ and $\mathcal{T}_2\left[\beta+\Omega_{2,1}/\Omega_m\right]$ include resonance terms in~\eqN{eq_EoM_2DoF_e1_freqs} with $j=2$. The removal of these four sets of resonance terms gives:
\begin{equation}
\dv{A_1}{\nu} = \frac{1-\cos{\phi}}{8 \Omega_{2,1} \Omega_m} B_2 + \frac{\sin{\phi}}{8 \Omega_{2,1} \Omega_m} A_2,\nonumber
\end{equation}\vspace{-3mm}
\begin{equation} 
\dv{B_1}{\nu} = -\frac{1-\cos{\phi}}{8 \Omega_{2,1} \Omega_m} A_2 + \frac{\sin{\phi}}{8 \Omega_{2,1} \Omega_m} B_2,\nonumber
\end{equation}\vspace{-3mm}
\begin{equation}
\dv{A_2}{\nu} = \frac{1-\cos{\phi}}{8 \Omega_{2,2} \Omega_m} B_1 - \frac{\sin{\phi}}{8 \Omega_{2,2} \Omega_m} A_1,\nonumber
\end{equation}\vspace{-3mm}
\begin{equation} 
\dv{B_2}{\nu} = \frac{-1+\cos{\phi}}{8 \Omega_{2,2} \Omega_m} A_1 - \frac{\sin{\phi}}{8 \Omega_{2,2} \Omega_m} B_1,\nonumber
\end{equation}
and 
\begin{equation}
\label{eq_22}
\dv[2]{A_j}{\nu} = \frac{\cos{\phi}-1}{32 \Omega_{2,1} \Omega_{2,2} \Omega_m^2} A_j, \dv[2]{B_j}{\nu} = \frac{\cos{\phi}-1}{32 \Omega_{2,1} \Omega_{2,2} \Omega_m^2} B_j.
\end{equation}
Modulation frequency $\Omega_m=(\Omega_{2,2}-\Omega_{2,1})/\beta$ is not a UMF in this case because $\cos{\phi}-1 \leq 0$. 

The results of the perturbation analysis of UMFs for the 2-DoF system are summarized in Table~\ref{tb_02}. 
\begingroup
\setlength{\tabcolsep}{6pt} 
\renewcommand{\arraystretch}{1} 
\begin{ruledtabular}
\begin{table}[htb]
\caption{Summary of the perturbation analysis of UMF for the 2-DoF system. 
\label{tb_02}}
{\centering
\begin{tabular}{l c c}
\multirow{2}{3em}{$\Omega_m$} & \multirow{2}{10em}{$\displaystyle \dv[2]{A_j}{\nu}\!\Big/\!A_j \!=\! \dv[2]{B_j}{\nu}\!\Big/\!B_j$} & When is $\Omega_m$ \\
 & & a UMF?\vspace{0.6mm}\\
\hline \\[-0.9 em]
$2\Omega_{2,1}/\beta$& $\displaystyle \frac{1+\cos{\phi}}{32\Omega_{2,1}^2\Omega_m^2}$ & $\cos{\phi}\neq-1$\vspace{1mm}\\
$2\Omega_{2,2}/\beta$& $\displaystyle \frac{1+\cos{\phi}}{32\Omega_{2,2}^2\Omega_m^2}$ & $\cos{\phi}\neq-1$\vspace{1mm}\\
$(\Omega_{2,1}\!+\!\Omega_{2,2})/\beta$& $\displaystyle \frac{1-\cos{\phi}}{32\Omega_{2,1}\Omega_{2,2}\Omega_m^2}$ & $\cos{\phi}\neq1$\vspace{1mm}\\
$(\Omega_{2,2}\!-\!\Omega_{2,1})/\beta$& $\displaystyle \frac{\cos{\phi}-1}{32\Omega_{2,1}\Omega_{2,2}\Omega_m^2}$ & Never.
\end{tabular}\par }
\end{table}
\end{ruledtabular}
\endgroup

\subsubsection{3-DoF modulated systems}
\label{sec:3DoF}

The natural frequencies of the unmodulated 3-DoF system are $\Omega_{3,1}=1$, $\Omega_{3,2}=\sqrt{1+K_c}$ and $\Omega_{3,3}=\sqrt{1+3K_c}$. Using the same procedure outlined in Section~\ref{sec:2DoF}, we can obtain expressions for the UMFs of the system by removing the terms that lead to exponential growth of the response. The results of this analysis are summarized in Table~\ref{tb_0}. We note that the 3-DoF system has a significantly larger number of UMFs than the 2-DoF system. 
\begingroup
\setlength{\tabcolsep}{6pt} 
\renewcommand{\arraystretch}{1} 
\begin{ruledtabular}
\begin{table}[htb]
\caption{Summary of the perturbation analysis of UMF for the 3-DoF system. 
\label{tb_0}}
{\centering
\begin{tabular}{l c c}
\multirow{2}{3em}{$\Omega_m$} & \multirow{2}{10em}{$\displaystyle \dv[2]{A_j}{\nu}\!\Big/\!A_j \!=\! \dv[2]{B_j}{\nu}\!\Big/\!B_j$} & When is $\Omega_m$ \\
 & & a UMF?\vspace{0.6mm}\\
\hline \\[-0.9 em]
$2\Omega_{3,1}/\beta$& $\displaystyle \frac{(1+2\cos{\phi})^2}{144\Omega_{3,1}^2\Omega_m^2}$ & $\cos{\phi}\neq-0.5$\vspace{1mm}\\
$2\Omega_{3,2}/\beta$& $\displaystyle \frac{\cos^2{\phi}}{16\Omega_{3,2}^2\Omega_m^2}$ & $\cos{\phi}\neq0$\vspace{1mm}\\
$2\Omega_{3,3}/\beta$& $\displaystyle \frac{(2+\cos{\phi})^2}{144\Omega_{3,3}^2\Omega_m^2}$ & Always.\vspace{1mm}\\
$(\Omega_{3,1}\!+\!\Omega_{3,2})/\beta$& $\displaystyle \frac{\sin^2{\phi}}{24\Omega_{3,1}\Omega_{3,2}\Omega_m^2}$ & $\sin{\phi}\neq0$\vspace{1mm}\\
$(\Omega_{3,2}\!-\!\Omega_{3,1})/\beta$& $\displaystyle \frac{-\sin^2{\phi}}{24\Omega_{3,1}\Omega_{3,2}\Omega_m^2}$ & Never.\vspace{1mm}\\
$(\Omega_{3,1}\!+\!\Omega_{3,3})/\beta$& $\displaystyle \frac{(1-\cos{\phi})^2}{72\Omega_{3,1}\Omega_{3,3}\Omega_m^2}$ & $\cos{\phi}\neq1$\vspace{1mm}\\
$(\Omega_{3,3}\!-\!\Omega_{3,1})/\beta$& $\displaystyle \frac{-(1-\cos{\phi})^2}{72\Omega_{3,1}\Omega_{3,3}\Omega_m^2}$ & Never.\vspace{1mm}\\
$(\Omega_{3,2}\!+\!\Omega_{3,3})/\beta$& $\displaystyle \frac{\sin^2{\phi}}{48\Omega_{3,2}\Omega_{3,3}\Omega_m^2}$ & $\sin{\phi}\neq0$\vspace{1mm}\\
$(\Omega_{3,3}\!-\!\Omega_{3,2})/\beta$& $\displaystyle \frac{-\sin^2{\phi}}{48\Omega_{3,2}\Omega_{3,3}\Omega_m^2}$ & Never.
\end{tabular}\par }
\end{table}
\end{ruledtabular}
\endgroup

\subsubsection{$n$-DoF modulated systems with $n\geq4$}
\label{sec:nDoF}

Using the same procedure for modulated systems with $n \geq4$, we can obtain a general expression for the UMFs as $\left( \Omega_{n,j_1}+\Omega_{n,j_2} \right)/\beta$, where $j_{1,2} \in \{1,2,\cdots,n\}$. 

The natural frequencies of the unmodulated system are bounded within $1\le\Omega_{n,j_{1,2}}<\sqrt{1 + 4K_c}$ for any value of $n$ because of the periodicity of the system. Because the general expression for UMFs in our perturbation analysis involves a combination of two unmodulated frequencies, we can conclude that the UMFs are bounded within $0<\Omega_m<2\sqrt{1 + 4K_c}$. 

\section{Stability diagrams for short systems} \label{sec_short}

We investigate the parametric instability of short systems that have well separated modes ($K_c=0.6$). We explore the influence of modulation amplitude, $K_m$, on stability in the range of modulation frequency, $0.1\le\Omega_m\le4$. This frequency range is chosen to include all the UMFs for $K_c=0.6$ and $\beta \leq20$, as discussed in Section~\ref{sec:nDoF}. To present the stability diagrams clearly, this frequency range is split at $\Omega_m=1.8$ and presented separately. The parametric instabilities occurring within $1.8\leq \Omega_m \leq4$ correspond to $\beta=1$, while those occurring within $0.1\leq \Omega_m \leq1.8$ correspond to $\beta \geq2$. We present the latter in logarithmic scale to improve clarity. 
In this study, all stability diagrams are presented at a resolution of 300 dots per inch (dpi). Consequently, the limitations associated with this resolution preclude the representation of certain minute features of stability diagrams in undamped systems. 

\subsection{$1.8 \leq \Omega_m \leq 4$}

\begin{figure}[htb]
\centerline{\includegraphics[scale=0.45]{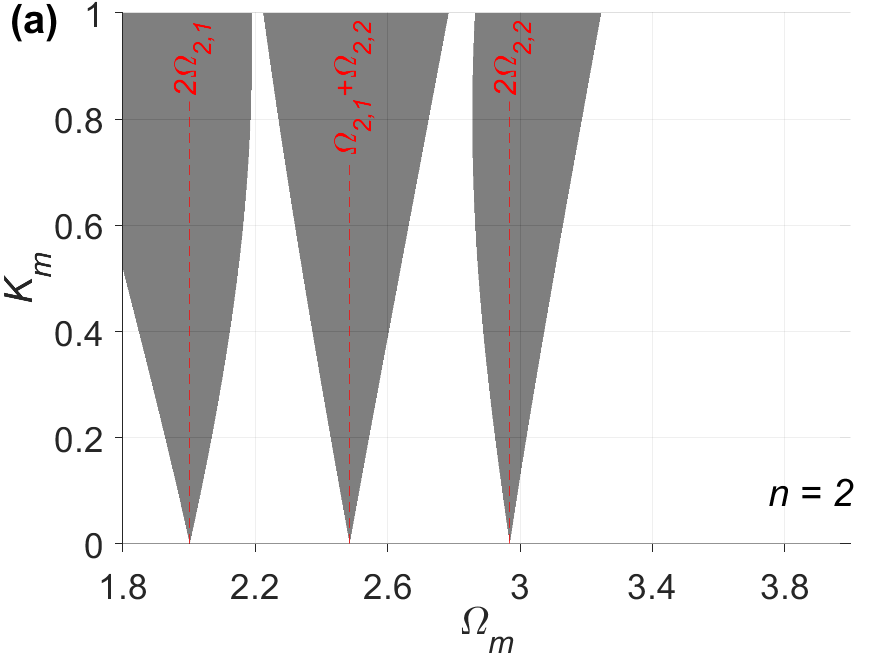}}
\centerline{\includegraphics[scale=0.45]{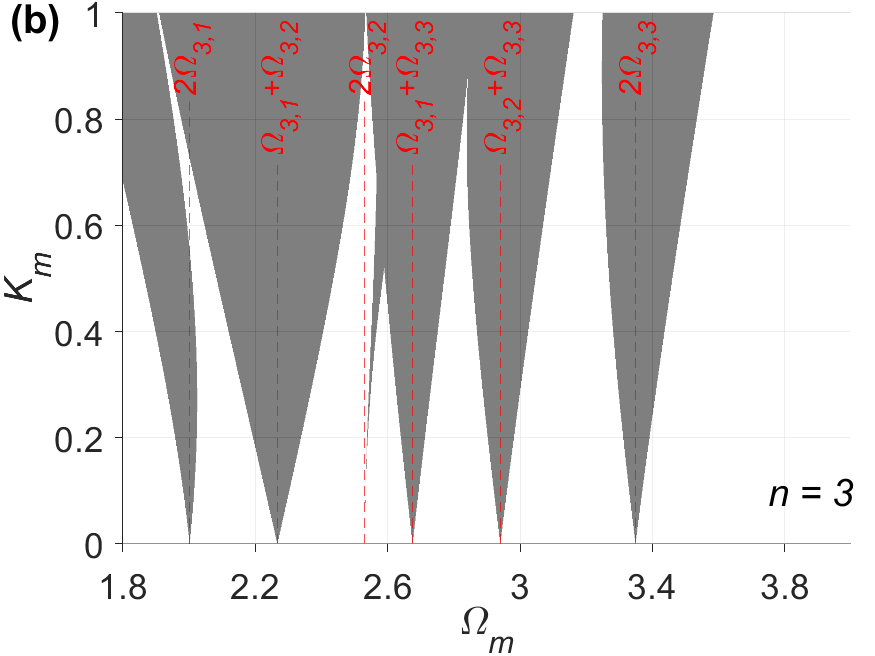}}
\centerline{\includegraphics[scale=0.45]{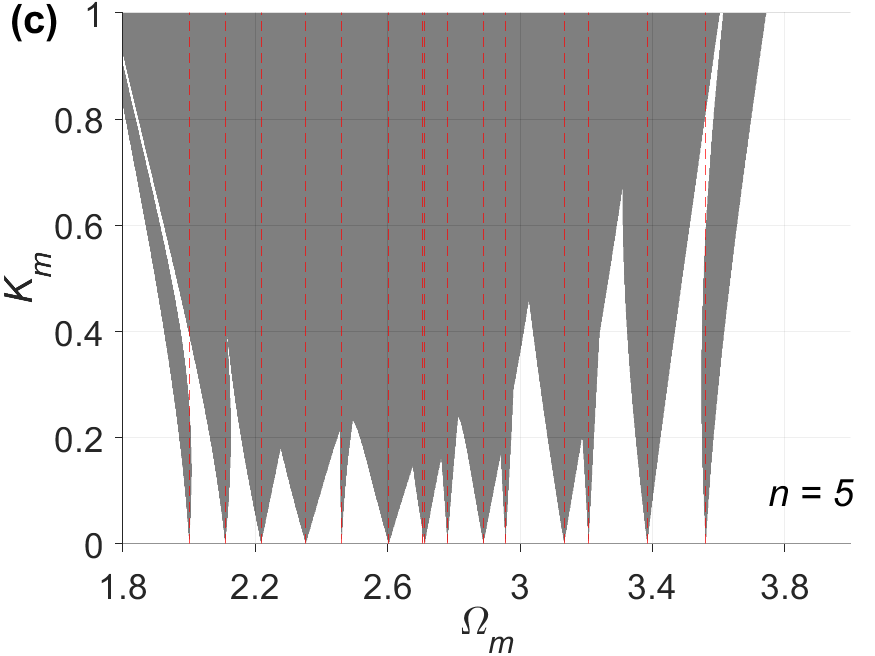}}
\caption{Stability diagrams for $K_c=0.6$, $\phi=0.5\pi$, $\zeta=0$ and different numbers of modulated units: (a) $n=2$, (b) $n=3$ and (c) $n=5$. Grey regions represent unstable response, white regions represent stable response. Red dashed lines indicate UMFs obtained from perturbation analysis.}\label{fig_STB_1}
\end{figure}

Fig.~\ref{fig_STB_1} shows the stability diagrams in the $(\Omega_m,K_m)$ plane for $1.8\leq \Omega_m \leq4$. The grey regions represent the combinations of $\Omega_m$ and $K_m$ that result in unstable response, while the white regions represent stable response. 
The unstable regions originate from the UMFs along $K_m=0$. The widths of these unstable regions increase as $K_m$ increases, giving the unstable regions an inverted triangular shape known as a {\it tongue}. 
The vertical dashed lines indicate the UMFs obtained from the perturbation analysis in Section~\ref{sec:perturbation}. We observe an excellent match between the predicted and computed UMFs. 

Many of the unstable regions have a triangular shape for small values of $K_m$. In the vicinity of the UMFs (near $K_m=0$), the slopes of the transition curves are generally symmetric with respect to the vertical line. In some cases, however, the transition curves become very close to each other and appear to touch the $K_m$-axis vertically; {\it e.g.} near $\Omega_m=2.53$ in Fig.~\ref{fig_STB_1}(b) and near $\Omega_m=2.46$ in Fig.~\ref{fig_STB_1}(c). These points correspond to the special values of $\phi$ at which the expression for the UMFs did not hold in the perturbation analysis; recall the exclusion points in the right-most column in Tables~1 and 2. At higher values of modulation amplitude, typically $K_m>0.2$, the transition curves bend and some adjacent unstable regions merge. 

We note that it is possible to obtain the slopes (and higher-order approximations) of the transition curves using the perturbation method outlined in Section~\ref{sec:perturbation}. This analysis yields good approximation of the transition curves, but is tractable only for $\beta=1$ and very small values of $n$. For example, we found the perturbation method to give the same slope for the transition curves as those reported in Figs.~\ref{fig_STB_1} (a) and (b), including the vertical tangency at $\Omega_m=2\Omega_{3,2}$ for $n=3$. We have not presented these calculations here because they are not the main focus in this work.

\subsection{$0.1 \leq \Omega_m \leq 1.8$} 
\label{sec_4.2}

\begin{figure}[htb]
\centerline{\includegraphics[scale=0.45]{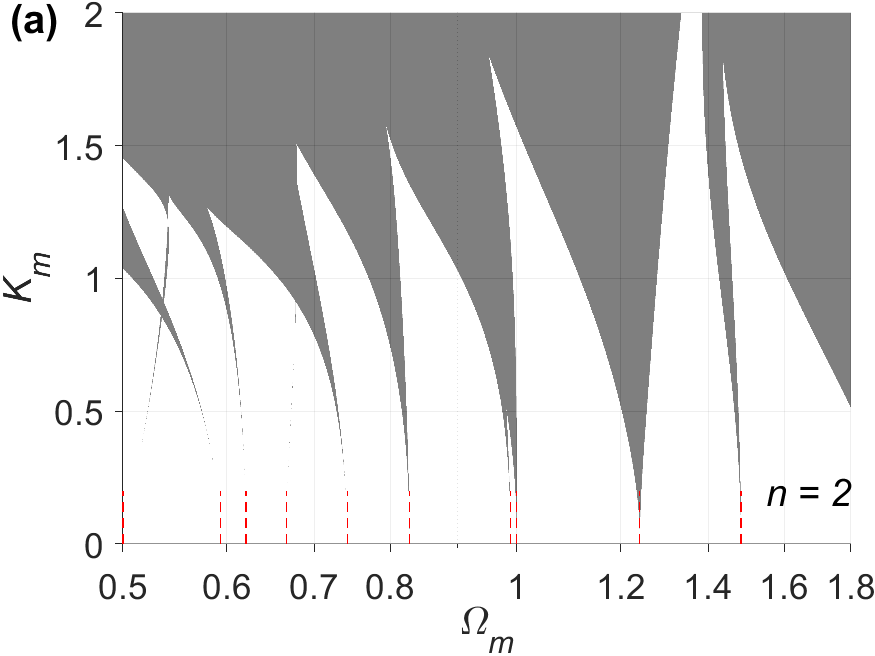}}
\centerline{\includegraphics[scale=0.45]{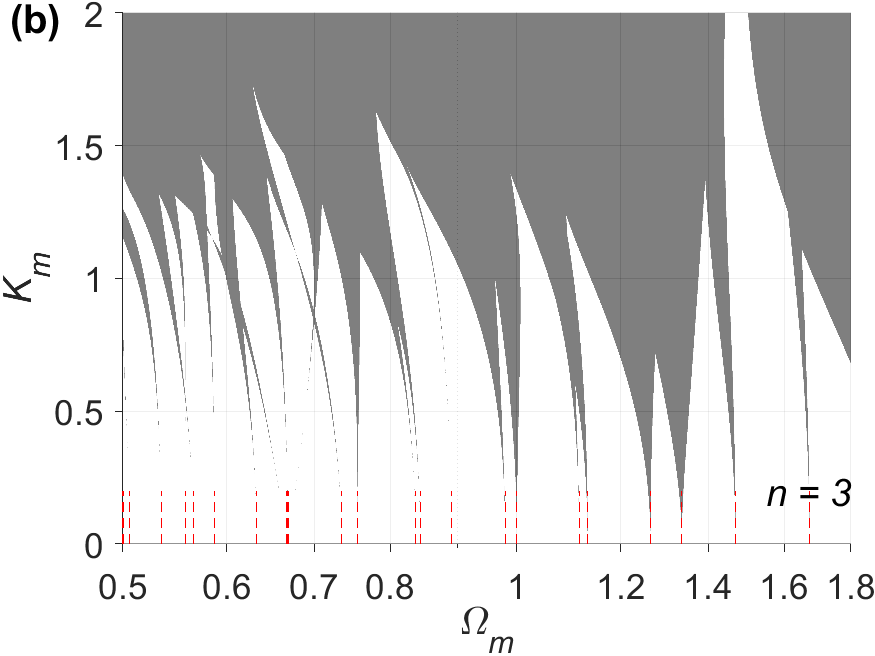}}
\centerline{\includegraphics[scale=0.45]{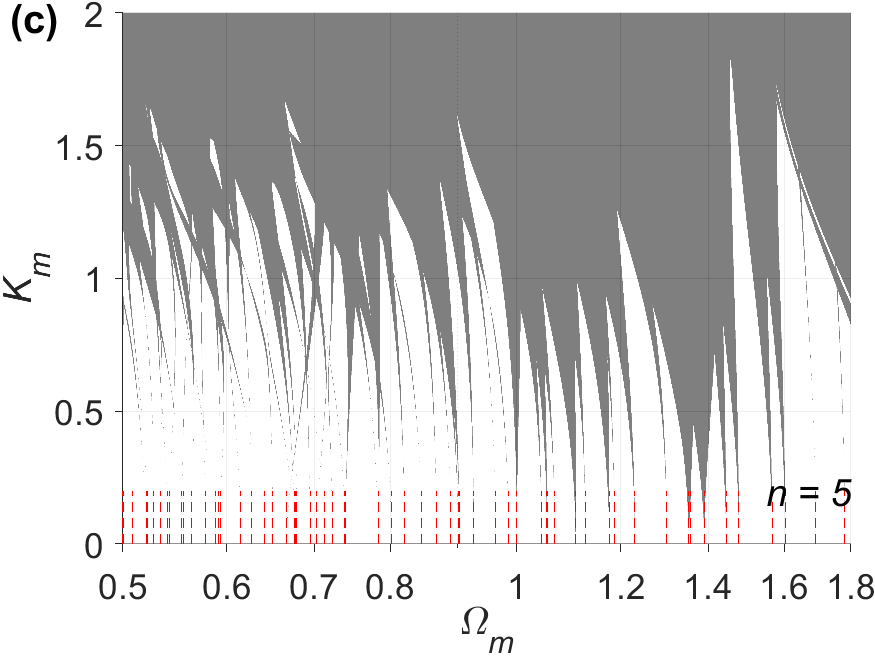}}
\caption{Stability diagrams for $K_c=0.6$ and $\phi=0.5\pi$. (a) $n=2$, (b) $n=3$ and (c) $n=5$. Red dashed lines indicate UMFs with $\beta \geq 2$, predicted by perturbation analysis.}
\label{fig_STB_2}
\end{figure}

\begin{figure}[htb]
\centerline{\includegraphics[scale=0.45]{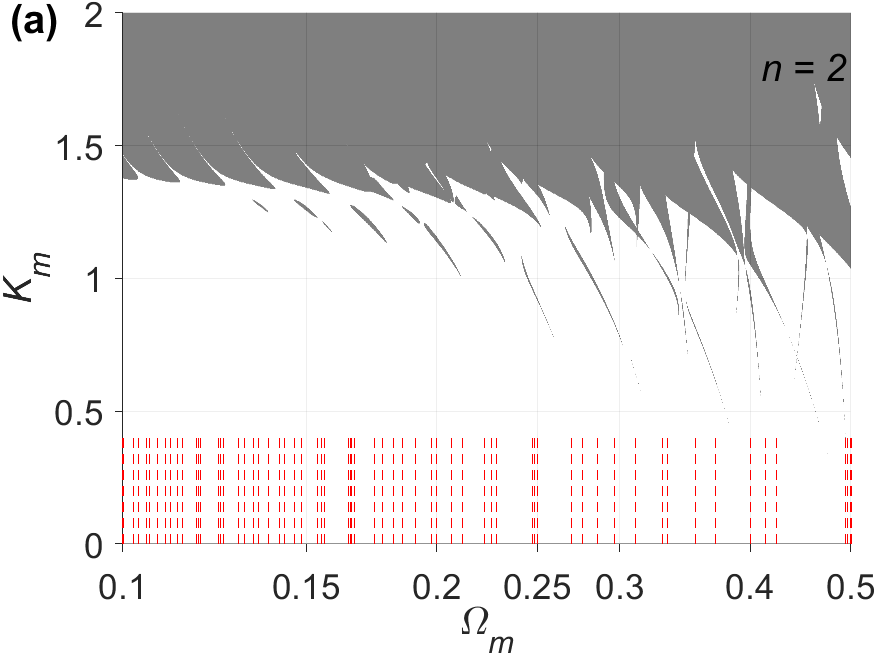}}
\centerline{\includegraphics[scale=0.45]{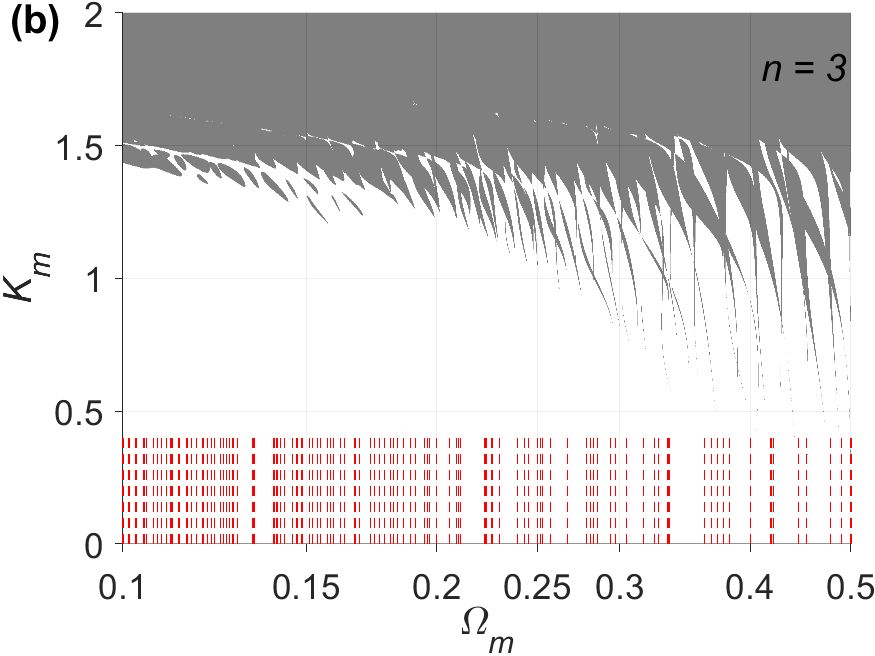}}
\centerline{\includegraphics[scale=0.45]{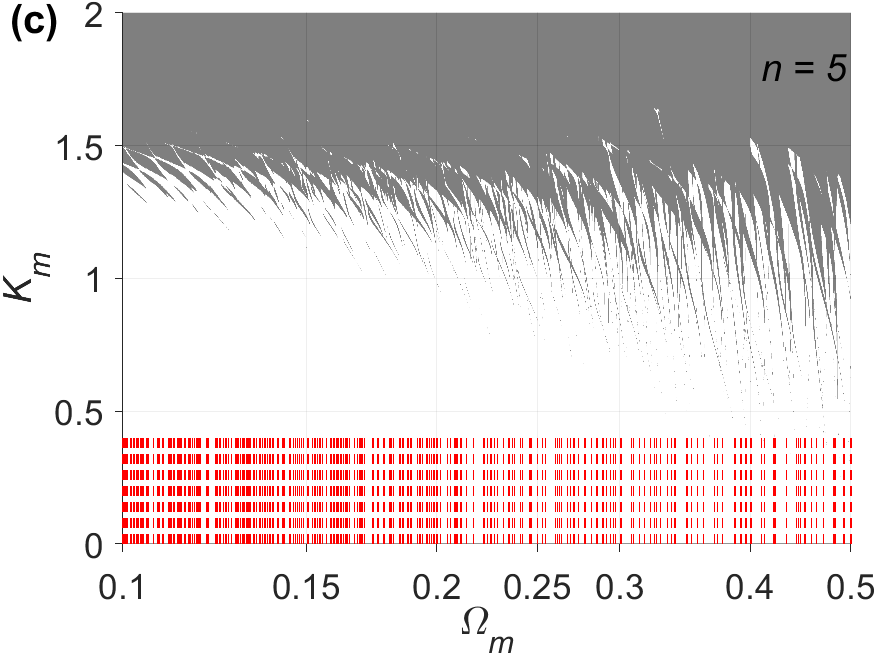}}
\caption{Stability diagrams for $K_c=0.6$ and $\phi=0.5\pi$. (a) $n=2$, (b) $n=3$ and (c) $n=5$. Red dashed lines indicate the UMFs predicted by perturbation analysis.
}
\label{fig_STB_3}
\end{figure}

Figs.~\ref{fig_STB_2} and~\ref{fig_STB_3} show the stability diagrams of the modulated systems over $0.5 \leq \Omega_m \leq 1.8$ and $0.1 \leq \Omega_m \leq 0.5$, respectively. All other parameters remain the same as those in Fig.~\ref{fig_STB_1}. 
Within $0.1\le \Omega_m \le1.8$, which corresponds to $\beta>1$, the lowest point of each unstable region is generally found to be at a modulation amplitude greater than $K_m=0.05$. Moreover, the onset of parametric instability tends to occur at a higher value of $K_m$ as $\Omega_m$ decreases. The UMFs with $\beta > 1$ cannot predict the $\Omega_m$ value of the lowest point of an unstable region for $\Omega_m<0.8$. 

We observe that certain unstable regions narrow as $K_m$ increases, leading to points where unstable regions converge or pinch off. This phenomenon is shown in Fig.\ref{fig_STB_2}(a) near $\Omega_m=0.54$ and in Fig.\ref{fig_STB_2}(b) near $\Omega_m=0.58$ and $\Omega_m=0.68$. At these points, the unstable regions taper and may eventually separate, forming distinct upper and lower segments. 

In Fig.~\ref{fig_STB_3}, numerous unstable regions appear as isolated regions, having separated from the unstable regions above them, and appearing as detached islands of instability. These detached unstable regions become numerous as the number of modulated units increases. Within the range of $0.1 \le \Omega_m \le 0.2$, the vertices of some unstable regions appear to be smooth (as opposed to pinched) at the resolution used for the computations. Given that many of these features disappear in the presence of damping (Section~\ref{sec_zeta}), we have not explored them in greater detail at present. 

We observe in Fig.~\ref{fig_STB_3} that large continuous regions of stable response emerge for smaller values of $\Omega_m$. For example, there are no unstable regions within the ranges $\ 0.1 \leq \Omega_m \leq 0.35,K_m \leq 0.5$ and $\ 0.1 \leq \Omega_m \leq 0.17,K_m \leq 1$. These stable regions provide relatively broad ranges for the safe operation of modulated devices with slow modulations. 

\section{The role of spatial modulation} 
\label{sec_phi}

\subsection{Stability diagrams in the $(\Omega_m,\phi)$ plane}

\begin{figure*}[htb]
\centerline{\includegraphics[scale=0.4]{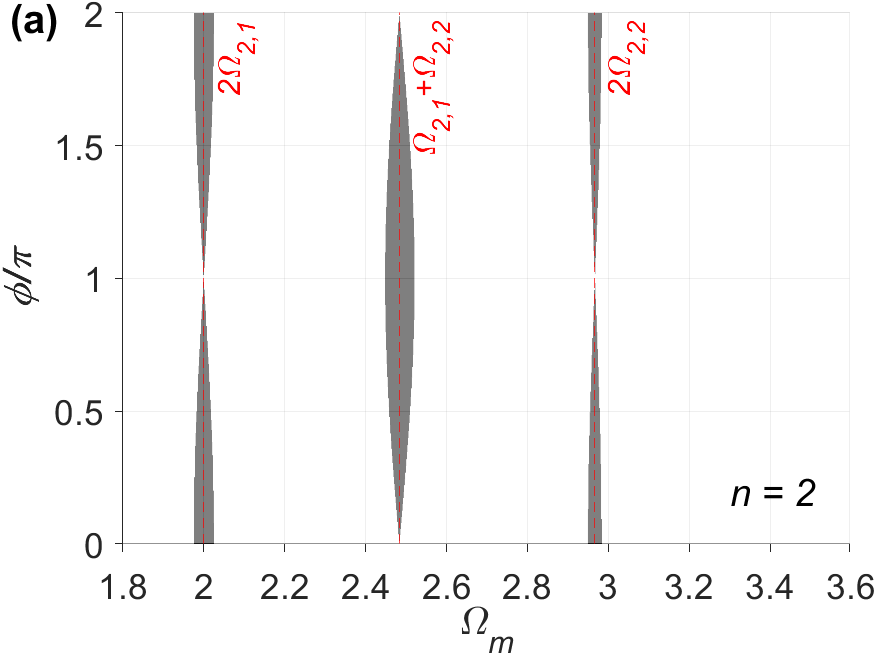}\includegraphics[scale=0.4]{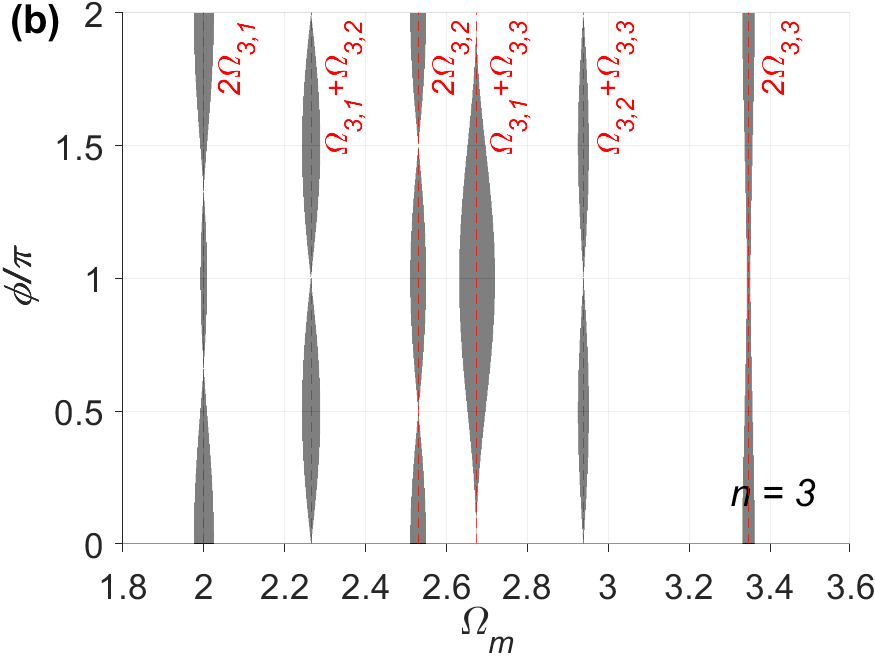}\includegraphics[scale=0.4]{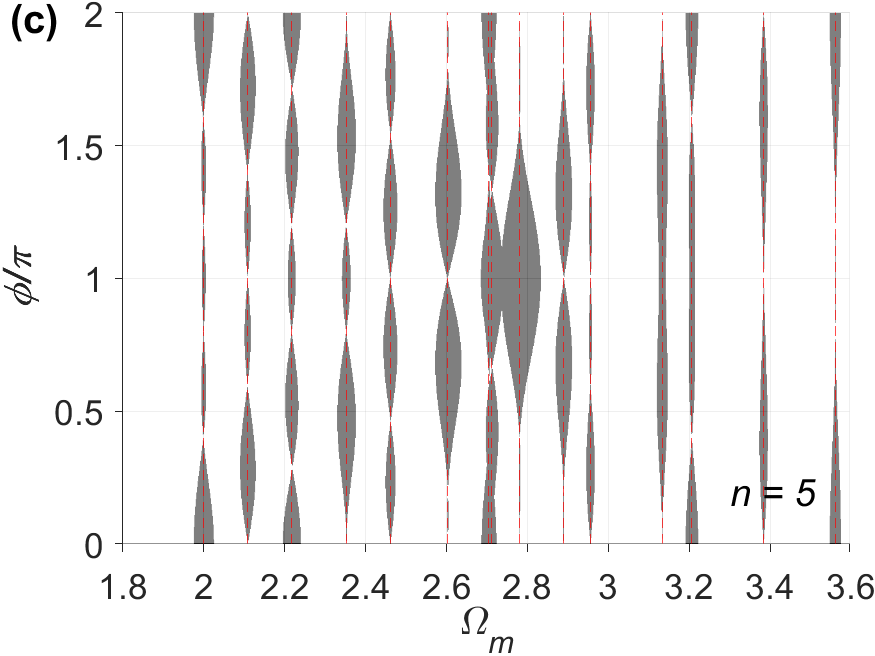}}
\caption{Stability diagrams for $K_c=0.6$ and $K_m=0.05$. (a) $n=2$, (b) $n=3$ and (c) $n=5$. Red dashed lines indicate UMFs predicted by perturbation analysis.}\label{fig_STB_4}
\end{figure*}

Fig.~\ref{fig_STB_4} shows the stability diagrams in the $(\Omega_m,\phi)$ plane for weakly modulated systems ($K_m=0.05$) and $1.8 \leq \Omega_m \leq 3.6$. This frequency range corresponds to resonance tongues with $\beta=1$. Most of the unstable regions do not overlap with each other, a feature that makes it easier to distinguish different features of the stability diagrams. All the stability diagrams are symmetric with respect to $\phi=\pi$ because the modulation term remains unchanged under the transformation $\phi\mapsto2\pi-\phi$. Moreover, we have $\partial\lambda/\partial\phi=0$ at $\phi=\pi$, where $\lambda$ represents the eigenvalue of $\underline{\underline{\mathsf{E}}}$ with the largest modulus. 

The unstable regions are organized around the UMFs predicted by the perturbation analysis, with their widths undulating symmetrically about the UMFs. The modulation phase at which the width of an unstable region is zero agrees well with the analysis in Section~\ref{sec:perturbation}. An example can be seen in panel (a) for the 2-DoF system: $2\Omega_{2,j}$ is an UMF except when $\cos\phi=-1$. 
These properties of unstable regions may not hold, however, when two adjacent regions overlap. See the overlapping regions near $\Omega_m=2.7$ in panel (c) for an example. 

\begin{figure}[htb]
\centerline{\includegraphics[scale=0.48]{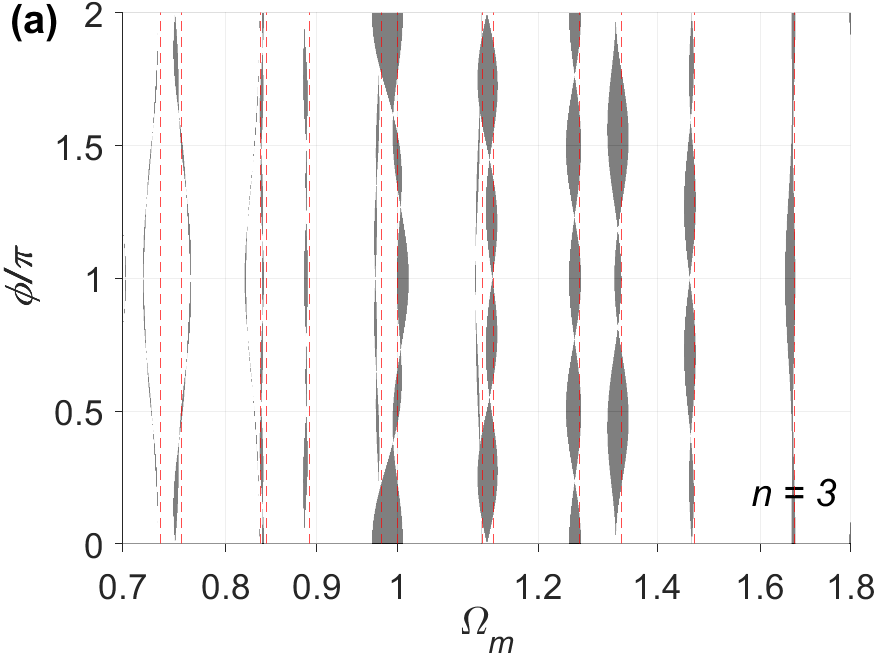}}
\centerline{\includegraphics[scale=0.48]{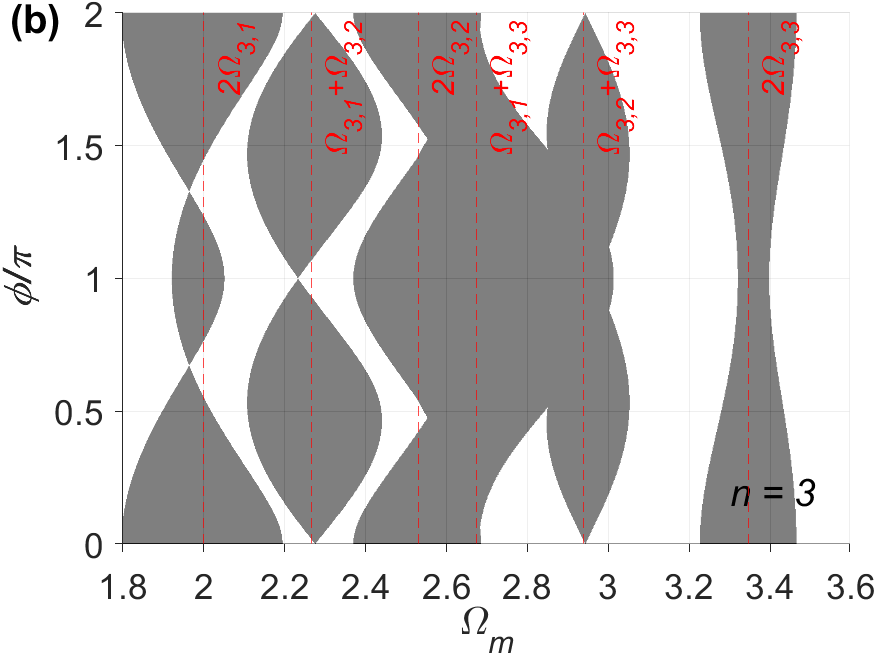}}
\caption{Stability diagrams for $K_c=0.6$, $K_m=0.4$ and $n=3$. (a) $0.7 \leq \Omega_m \leq 1.8$ for $\beta=1$, (b) $1.8 \leq \Omega_m \leq 3.6$ for $\beta>1$. Red dashed lines indicate UMFs predicted by perturbation analysis.
}
\label{fig_STB_5}
\end{figure}

Fig.~\ref{fig_STB_5} shows the stability diagrams in the $(\Omega_m,\phi)$ plane for a system with$K_m=0.4$ (strong modulation) and $n=3$. Even though the regions of stability are still organized around the UMFs, they are no longer symmetric about the UMFs; {\it cf.} Fig.~\ref{fig_STB_4}(b). The unstable regions become wider with increasing the modulation amplitude, as expected. The unstable regions that correspond to $2\Omega_{3,1}$, $\Omega_{3,1}+\Omega_{3,2}$ and $2\Omega_{3,3}$ show these features clearly. 
The same trend is observed for $\beta>1$ in panel (a). We also observe that the widths of the unstable regions in Fig.~\ref{fig_STB_5}(a) for $\beta \geq2$ alternate more frequently than those in Fig.~\ref{fig_STB_4}(b) for $\beta=1$. 

\subsection{Stability diagrams in the $(K_m,\phi)$ plane}

\begin{figure}[htb]
\centerline{\includegraphics[scale=0.4]{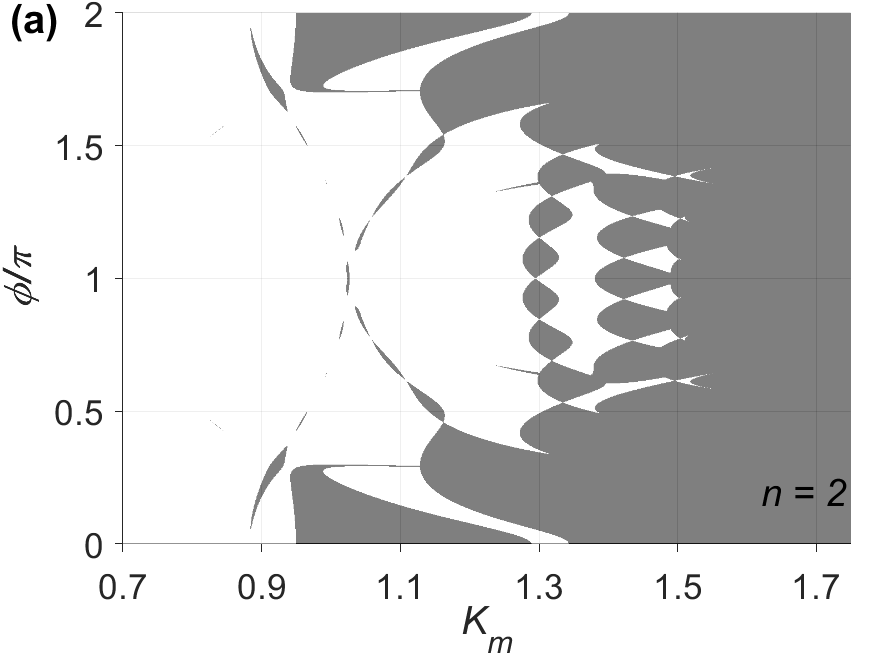}}
\centerline{\includegraphics[scale=0.4]{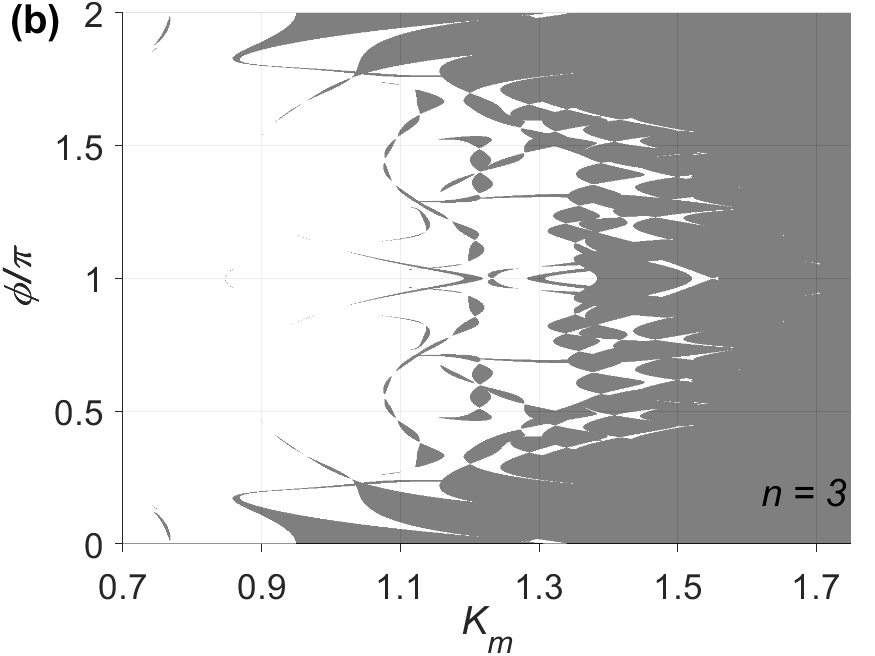}}
\centerline{\includegraphics[scale=0.4]{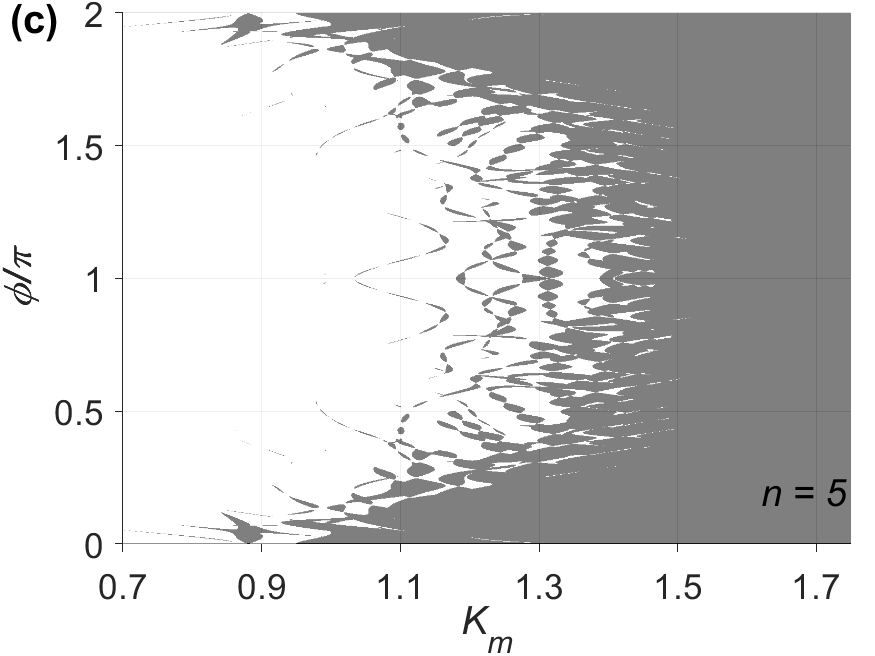}}
\caption{Stability diagrams for $K_c=0.6$ and $\Omega_m=0.2$. (a) $n=2$, (b) $n=3$ and (c) $n=5$. For all values of $\phi$, the response remains stable for $K_m\le0.7$ and unstable for $K_m\ge1.71$. 
}
\label{fig_STB_6}
\end{figure}

Fig.~\ref{fig_STB_6} shows the stability diagrams for three short systems with $K_c=0.6$ and $\Omega_m=0.2$. For all values of $\phi$, the response remains stable when $K_m\le0.7$ and unstable when $K_m\ge1.7$. Thus, all the transition curves are confined within the range $0.7<K_m<1.7$. 

The transition curves exhibit more complicated shapes when projected onto the $(K_m,\phi)$ plane. We observe for $n=2$, panel (a), the reappearance of stable regions that are surrounded by regions of instability. These islands of stable (unstable) response become more numerous, fragmented and elaborate as the number of units increases; see panels (b) and (c). Despite these intricacies, the range of modulation amplitude over which the response remains stable becomes wider as the modulation phase approaches $\pi$. 

\section{Stability of long systems} \label{sec_long}

\begin{figure}[htb]
\centerline{\includegraphics[scale=0.4]{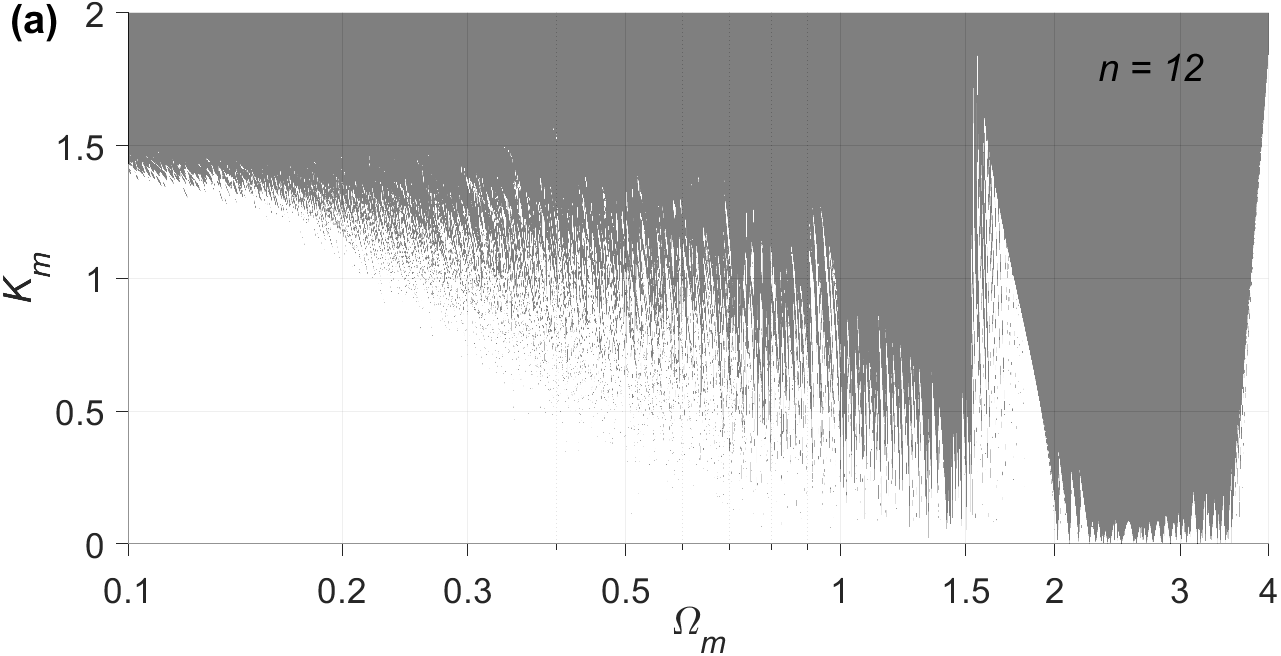}}
\centerline{\includegraphics[scale=0.4]{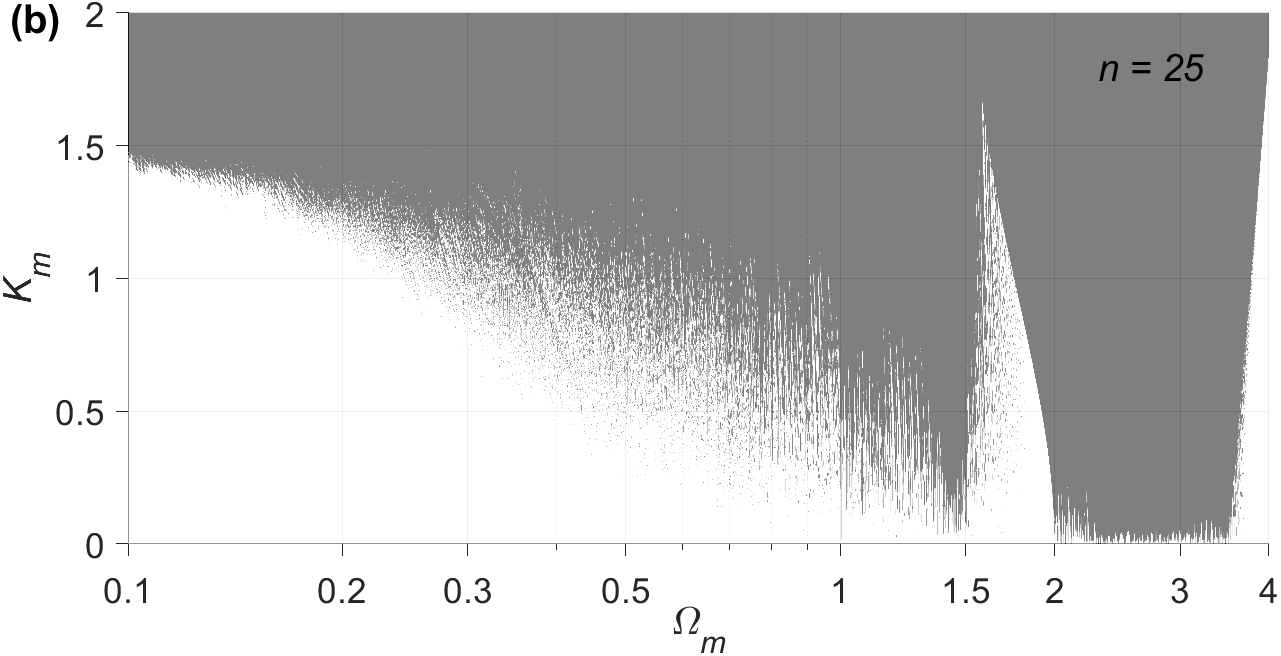}}
\centerline{\includegraphics[scale=0.4]{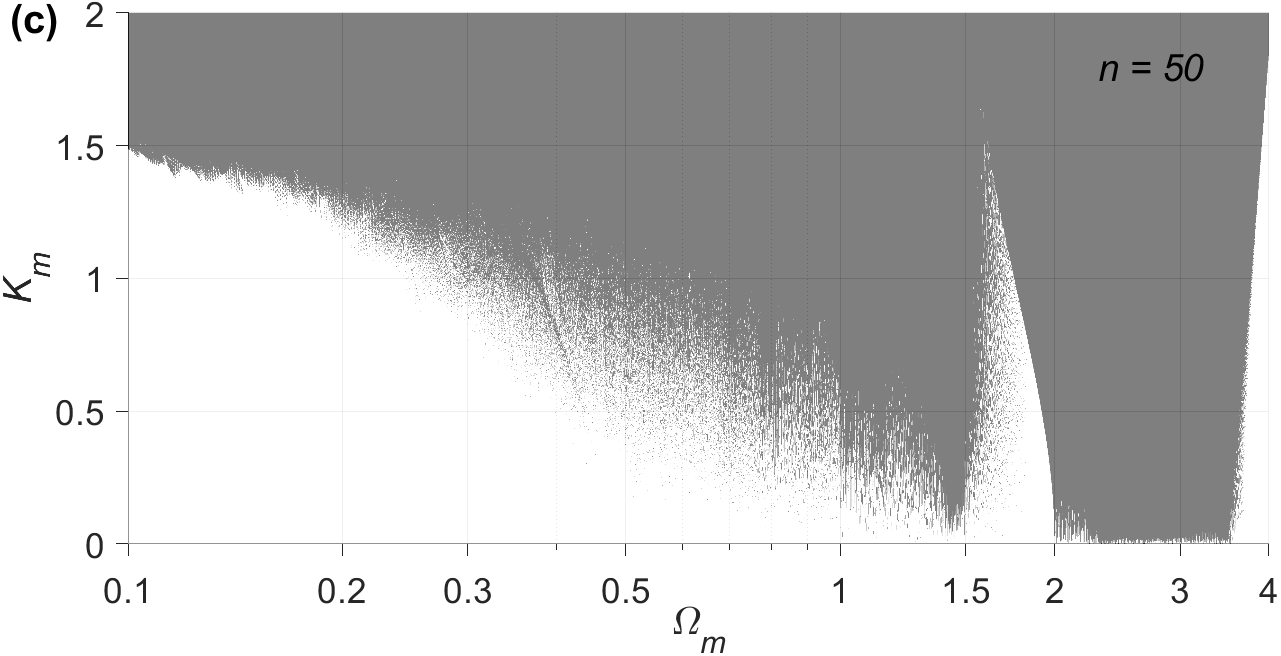}}
\caption{Stability diagrams of the modulated systems with $K_c=0.6$ and $\phi=0.5\pi$. (a) $n=12$, (b) $n=25$ and (c) $n=50$. The regions bounded by the pink curves are cloudy regions.}\label{fig_STB_7}
\end{figure}
For short systems ($n\le5$), Fig.~\ref{fig_STB_1} suggests that as the number of modulated units increases, there is a larger set of system parameters that leads to parametric instability: longer systems exhibit more unstable modulation frequencies for $\beta=1$, and the widths of the unstable regions increase with $K_m$. A similar trend cannot be seen easily for $\beta>1$: the stability diagrams in Figs.~\ref{fig_STB_2},~\ref{fig_STB_3} and~\ref{fig_STB_6} have an increasingly more complicated shape as $n$ increases, but it is no longer clear how the total areas covered by the unstable regions in these figures change. To further investigate the impact of $n$ on stability, we calculate the stability diagrams for long systems. 

Fig.~\ref{fig_STB_7} shows the stability diagrams in the $(\Omega_m,K_m)$ plane for three long systems with $K_c=0.6$ and $\phi=0.5\pi$. The unstable regions nearly cover the entire range of $2 \leq \Omega_m \leq 3.688$, within which the UMFs with $\beta=1$ are present. At lower modulation frequencies ($\beta>1$), where UMFs appear very densely, the unstable regions become fragmented and there is barely any clear transition curve in the range $0.1\leq \Omega_m \leq1.5$. 
For ease of reference, we refer to these regions as the {\it cloudy} regions of the stability diagrams. 

A main consistent trend in the cloudy regions is that as $\Omega_m$ decreases from $1.5$, the onset of instability occurs at a higher value of $K_m$. Many of the other features of the cloudy region do not seem to follow a clear pattern. There are several very small islands of stability that appear sporadically within the unstable regions at higher values of $K_m$. For $n=12$, Fig.~\ref{fig_STB_7}(a), two such stable regions appear above $K_m=1.5$ near $\Omega_m=0.4$. In contrast, the response remains unstable for $K_m\ge1.5$ for $n=25$. We have not explored these features in detail because they disappear in the presence of damping; see Section~\ref{sec_zeta}. 

Another prominent feature of the stability diagrams in Fig.~\ref{fig_STB_7} is the presence of a semi-triangular region within $1.5\leq \Omega_m \leq2$, the majority of which corresponds to stable response. This frequency range falls immediately below the lowest UMF with $\beta=1$. In these triangular regions, the system can exhibit stable response at very high values of $K_m$. Fragmented unstable regions are present within each triangular region, with greater density on the left side. For $0.1\leq \Omega_m \leq \sqrt{1+4K_c}$, the largest value of $K_m$ in a stable region occurs within the triangular region. 

\begin{figure}[htb]
\centerline{\includegraphics[scale=0.45]{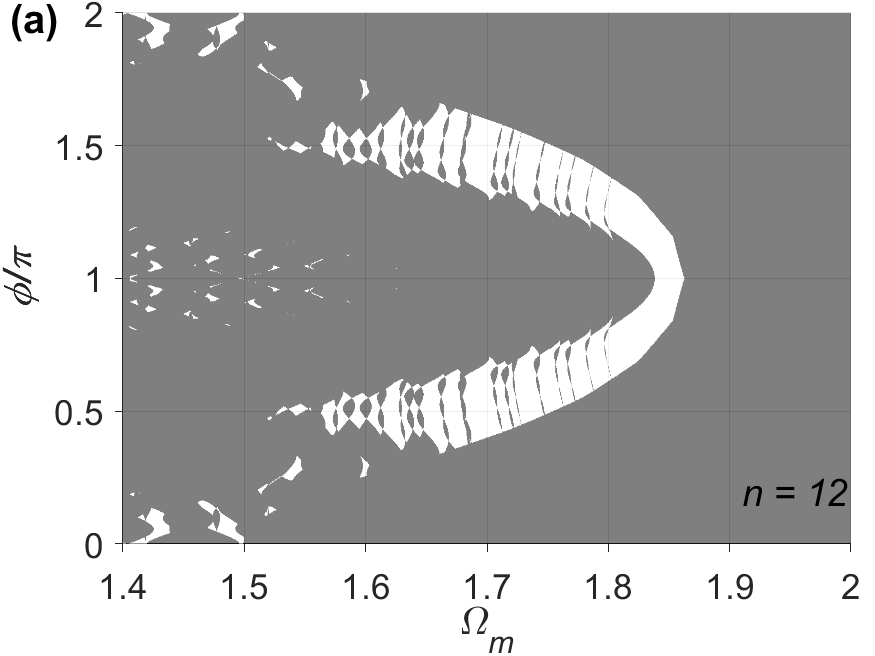}}
\centerline{\includegraphics[scale=0.45]{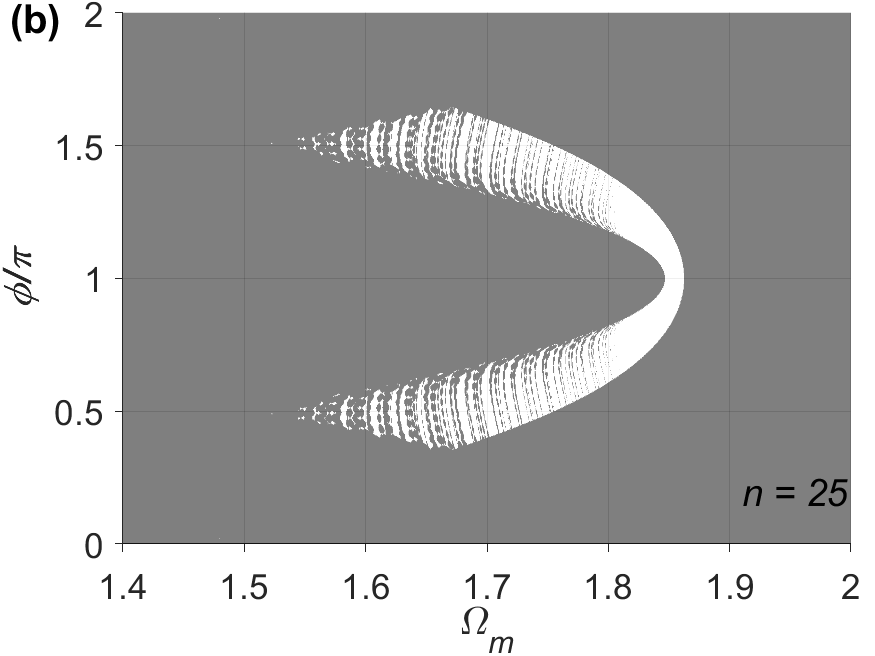}}
\centerline{\includegraphics[scale=0.45]{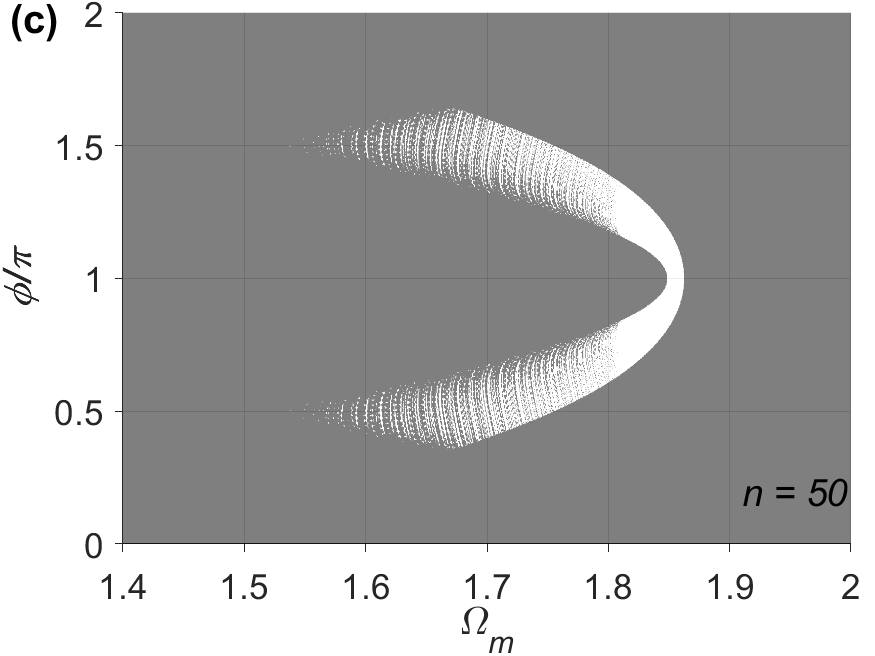}}
\caption{Stability diagrams for $K_c=0.6$ and $K_m=1$. (a) $n=12$, (b) $n=25$ and (c) $n=50$. }\label{fig_STB_8}
\end{figure}

To explore this phenomenon in greater detail, we calculate the stability diagrams in the $(\Omega_m,\phi)$ plane for $1.4\leq \Omega_m \leq2$. Fig.~\ref{fig_STB_8} shows the stability diagrams for three long systems with $K_c=0.6$ and $K_m=1$. A horizontal U-shaped region of predominantly stable response appears in each diagram within the range $0.35\pi\leq \phi \leq1.65\pi$ and $1.55\le\Omega_m\le1.81$. Several narrow bands of unstable response appear within the stable U-shaped regions. These band become increasingly more fragmented, narrow and dense as $n$ increases. For $\Omega_m \geq1.81$, no unstable regions are found within the U-shaped regions. We conjecture that the stability of this portion of the U-shaped region persists as the number of modulated units increases. We highlight the role of $\phi$ in the existence of such a persistence region of stable response in a strongly modulated system: the response at $\phi=0$ remains unstable. 

\begin{figure}[htb]
\centerline{\includegraphics[scale=0.45]{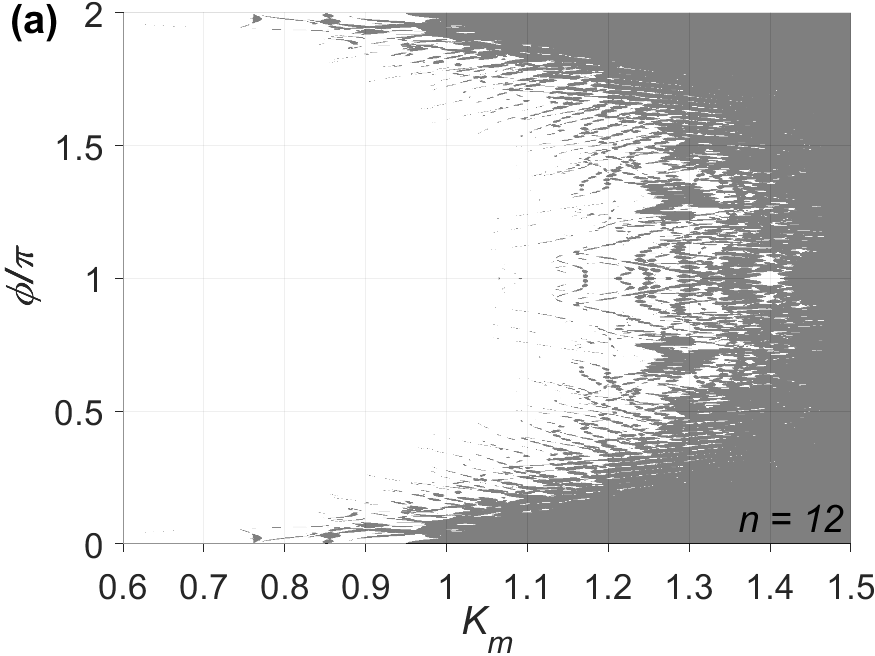}}
\centerline{\includegraphics[scale=0.45]{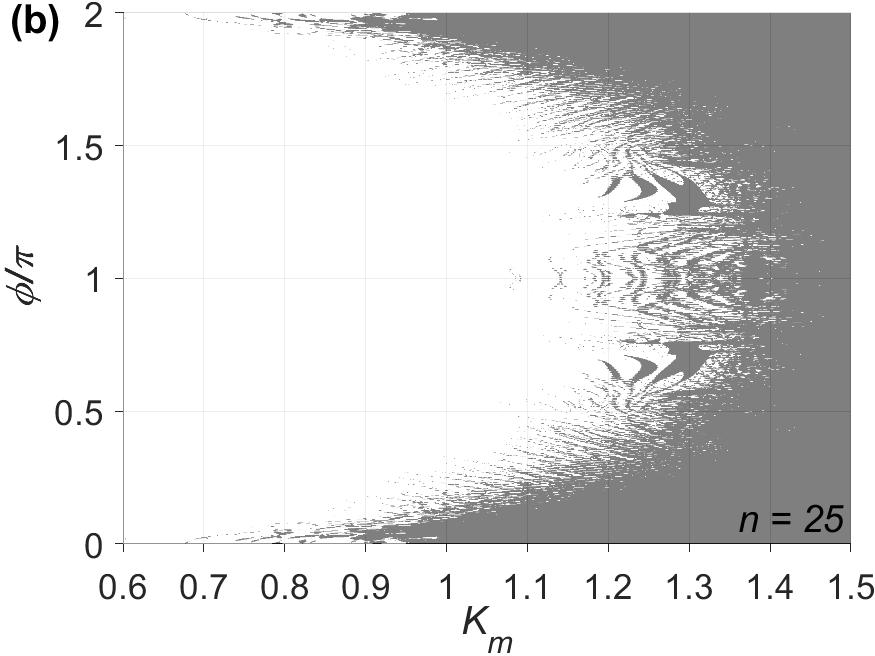}}
\centerline{\includegraphics[scale=0.45]{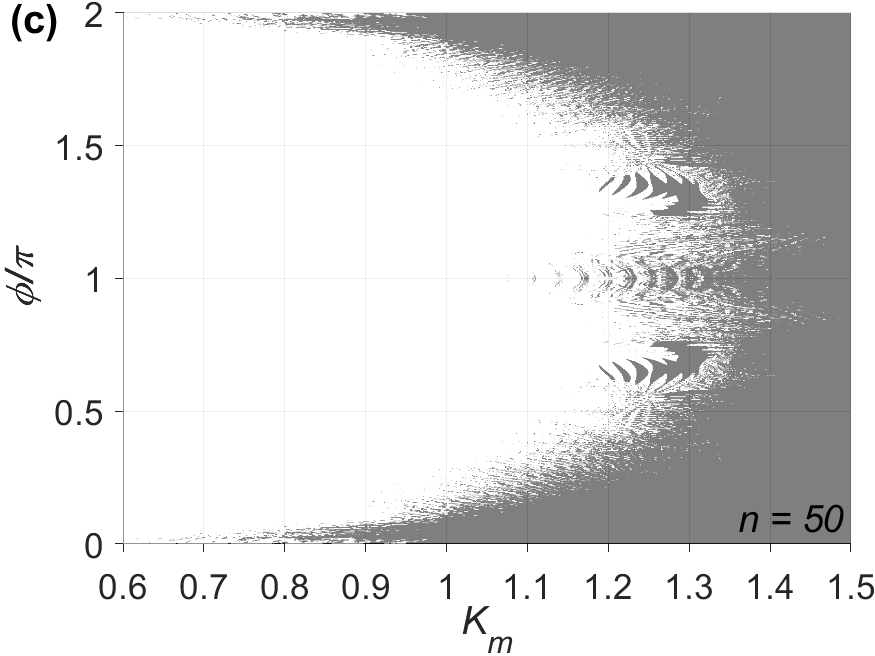}}
\caption{Stability diagrams for $K_c=0.6$ and $\Omega_m=0.2$. (a) $n=12$, (b) $n=25$ and (c) $n=50$. For all values of $\phi$, the response remains stable for $K_m\le0.6$ and unstable for $K_m\ge1.5$. }\label{fig_STB_9}
\end{figure}

Fig.~\ref{fig_STB_9} shows the stability diagrams in the $(\Omega_m,K_m)$ plane for three long systems with $K_c=0.6$ and $\Omega_m=0.2$. For these parameters, all transition curves are confined to the range $0.6<K_m<1.5$. In general, the range of modulation amplitudes over which the response remains stable becomes wider as the modulation phase increases from 0 to $\pi$. The fragmentation of the small regions of instability continues as $n$ increases, though it remains unclear whether the total area covered by the unstable regions increases or not. These are the same trends that we observed for shorter systems in Fig.~\ref{fig_STB_6}. The investigation fo the fine details of the fragmented transition between stable and unstable regions falls beyond the scope of this work. 

\section{Influence of damping}
\label{sec_zeta}

\begin{figure*}[htb]
\centerline{\includegraphics[scale=0.4]{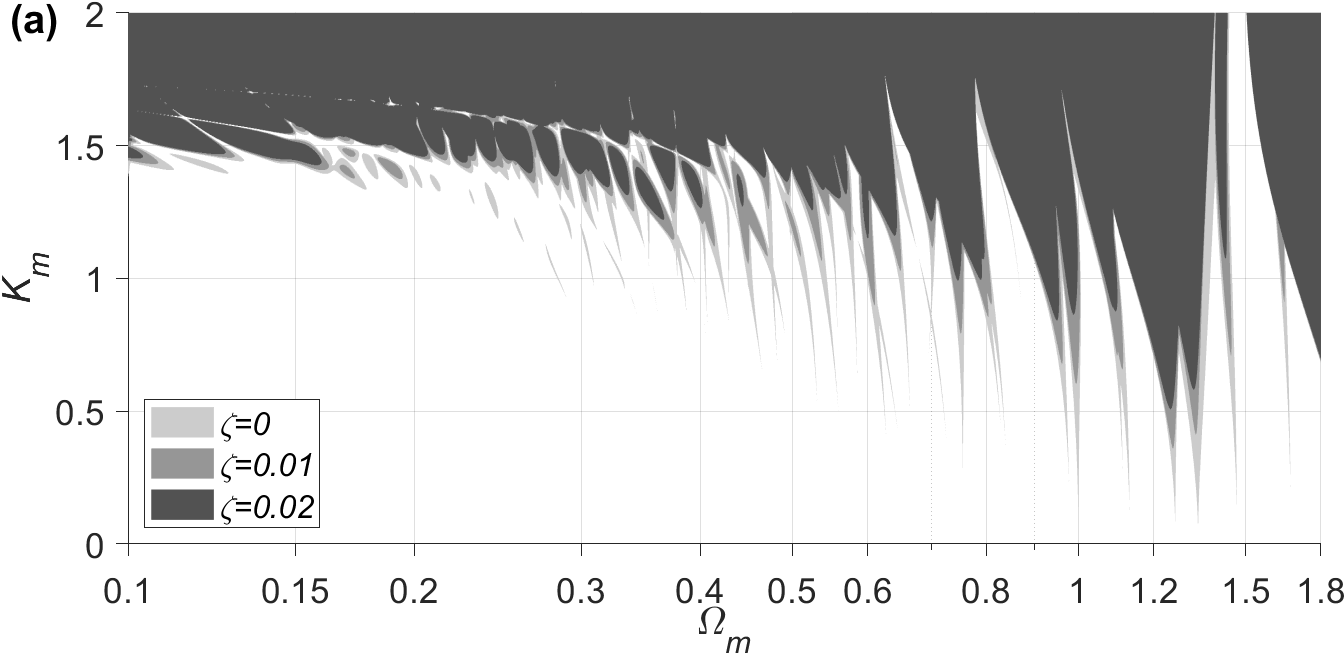}
\includegraphics[scale=0.4]{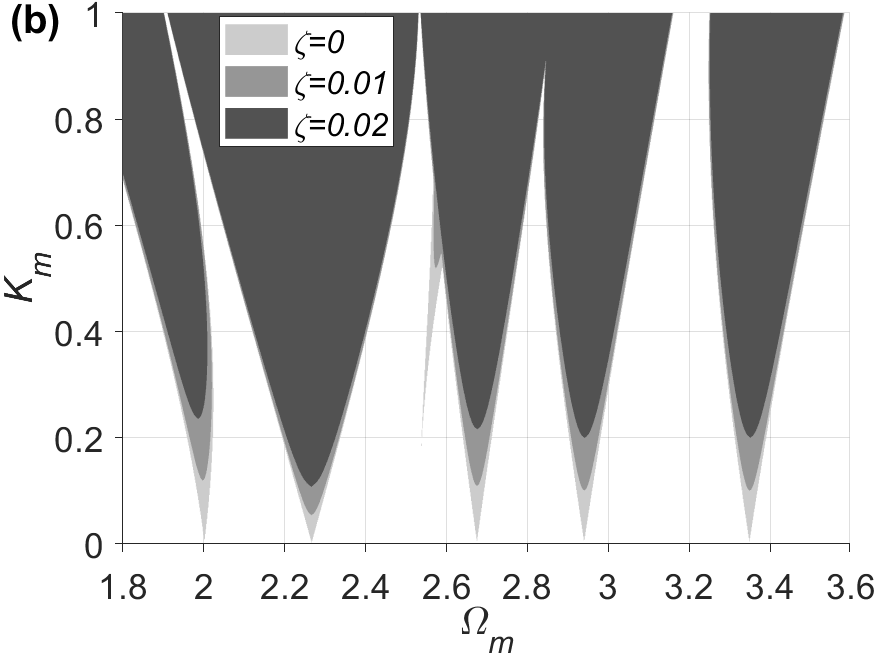}
}
\caption{Stability diagrams of the modulated systems with $n=3$, $K_c=0.6$ and $\phi=0.5\pi$. (a) $0.1\le\Omega_m\le1.8$; (b) $1.8\le\Omega_m\le3.6$. The darker grey zones indicate unstable regions corresponding to higher values of $\zeta$. }
\label{fig_STB_10}
\end{figure*}

\begin{figure*}[htb]
\centerline{\includegraphics[scale=0.4]{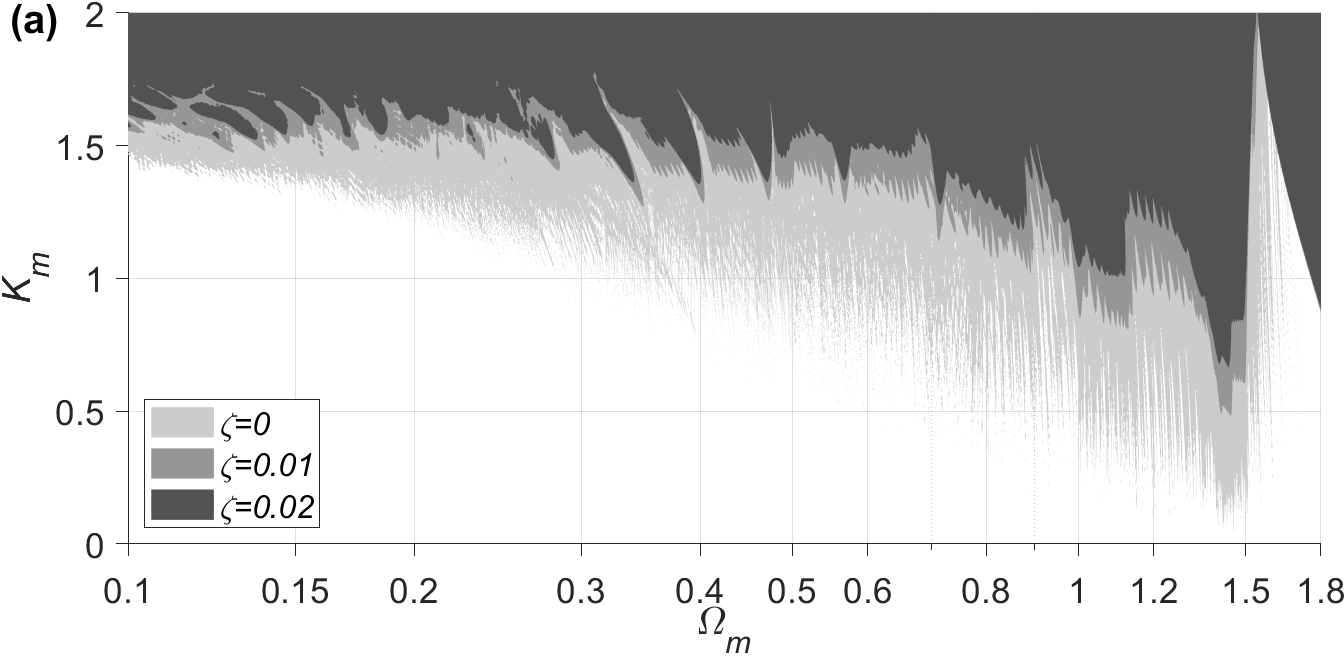}
\includegraphics[scale=0.4]{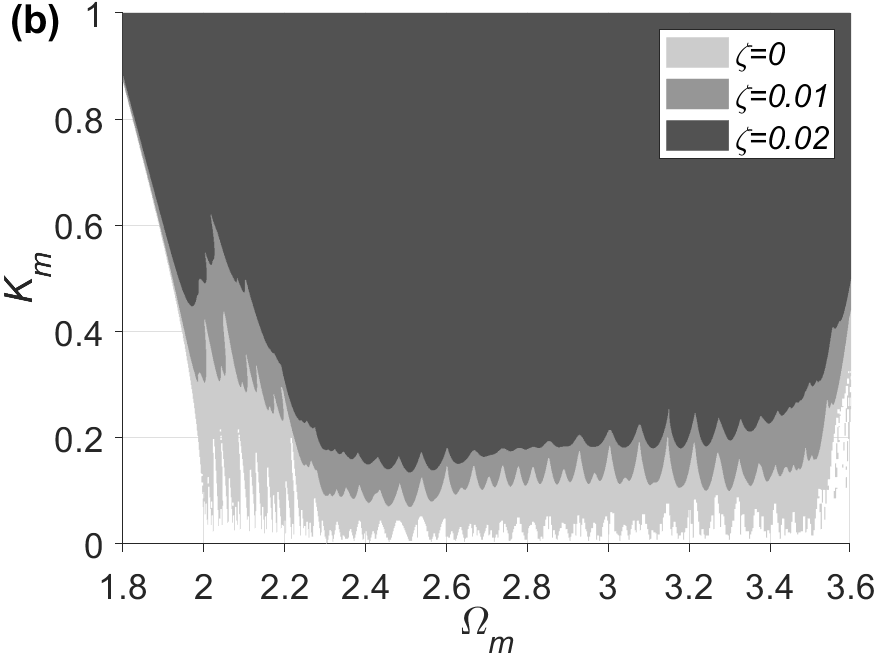}}
\caption{Stability diagrams of the modulated systems with $n=25$, $K_c=0.6$ and $\phi=0.5\pi$. (a) $0.1\le\Omega_m\le1.8$; (b) $1.8 \leq \Omega_m \leq 3.6$. The darker grey zones indicate unstable regions corresponding to higher values of $\zeta$. }
\label{fig_STB_11}
\end{figure*}
It is intuitively understood that damping enhances the stability of the response. To demonstrate this effect, Fig.~\ref{fig_STB_10} shows the stability diagrams for $n=3$, $K_c=0.6$ and different damping ratios. The darker grey zones indicate unstable regions corresponding to higher values of $\zeta$. Note that Fig.~\ref{fig_STB_10}(a) is plotted over the range $0\leq K_m \leq2$, while Fig.~\ref{fig_STB_10}(b) is plotted over $0\leq K_m \leq1$ for better clarity. 

As expected, the regions of unstable response become smaller as the damping ratio increases, with many of the smaller regions of instability disappearing; see Fig.~\ref{fig_STB_10}(a). The vertices of the tongues detach from the $K_m$ axis and become rounded, as seen in Fig.~\ref{fig_STB_10}(b). Thus, a minimum modulation amplitude is now required for parametric instability to occur. The transition curves for $\beta>1$ appear to be influenced by damping to a greater extent than those for $\beta=1$.  

The same trends are observed in the stability diagrams of a longer system ($n=25$) in Fig.~\ref{fig_STB_11}. In addition, panel (a) shows that the cloudy region of the stability diagrams has almost completely disappeared (become stable) in the damped system. The semi-triangular region of stable response in panel (a) becomes wider and extends to higher values of $K_m$ as $\zeta$ increases, with almost all the fragmented unstable regions within it disappearing. 

\begin{ruledtabular}
\begin{table}[htb]
\caption{Percentage of the unstable regions in the stability diagrams in Figs.~\ref{fig_STB_10} and~\ref{fig_STB_11}.\label{tb_1}}
{\centering
\begin{tabular}{p{.5in} p{0.6in} c c}
& & $0\leq K_m \leq2$ & $0\leq K_m \leq1$\\
& & $0.1\le\Omega_m \le1.8$ & $1.8\leq \Omega_m \leq3.6$\\
\hline
\multirow{3}{.5in}{$n=3$}& $\zeta=0$ & $46\%$ & $55\%$ \\
&$\zeta=0.01$ & $41\%$ & $53\%$ \\
&$\zeta=0.02$ & $38\%$ & $50\%$\\
\hline
\multirow{3}{.5in}{$n=25$}& $\zeta=0$ & $65\%$ & $91\%$ \\
&$\zeta=0.01$ & $40\%$ & $80\%$ \\
&$\zeta=0.02$ & $34\%$ & $72\%$\\
\end{tabular}\par }
\end{table}
\end{ruledtabular}


As a quantitative indication of the overall influence of damping on parametric instability, we calculate the percentage of the total area of the unstable regions in Figs.~\ref{fig_STB_10} and~\ref{fig_STB_11}; see Table~\ref{tb_1}. Although some of the stability diagrams use a logarithmic scale for $\Omega_m$, the areas are calculated using a linear scale.  
The introduction of damping results in a significant decrease in the unstable regions. The influence of damping is greater for $\beta>1$ than for $\beta=1$, as previously observed. In general, increasing $\zeta$ from 0 to 0.01 results in a greater change in the area than increasing it from 0.01 to 0.02. Moreover, the influence of damping is greater on the longer system. Thus, in damped systems with slow modulations, longer systems tend to provide greater stability compared to shorter ones. 

\section{Conclusions}
\label{sec_conclusions}

We presented a detailed computational investigation of parametric instability in a discrete model of a 1-D spatiotemporally modulated system. We assessed the stability of the response using direct computation based on Floquet theory. We explored the roles of several key parameters on parametric instability such as modulation parameters (modulation phase or wavenumber, amplitude, and frequency), damping and the number of units.

We used perturbation theory to show that unstable modulation frequencies (UMFs) occur at combinations of two natural frequencies of the underlying unmodulated system divided by a natural number, $\beta$. UMFs are modulation frequencies at which the response of the undamped system grows exponentially over time. Stability analysis based on Floquet theory confirmed that UMFs with $\beta = 1$ induce parametric instability in the undamped system regardless of how small the modulation amplitude is. In contrast, Floquet theory revealed that UMFs with $\beta \geq 2$ may not always lead to instability, provided that the modulation amplitude is sufficiently small. As the modulation frequency decreases, parametric instability occurs at increasingly higher modulation amplitudes. There are therefore significant regions of stable response within the stability diagram, offering a broad safe range of modulation amplitudes for designing stable modulated systems. 

As a function of the modulation phase (wavenumber), the unstable regions are symmetric with response to $\phi=\pi$. At low modulation amplitudes, the unstable regions appear symmetrically about the UMFs with $\beta = 1$. The perturbation analysis accurately predicts the values of modulation phase at which the width of an unstable region is zero (typically occurs multiple times). As the modulation amplitude increases, the unstable regions become wider, are no longer symmetric with respect to the UMFs, and overlap with each other. Overall, the range of modulation amplitudes over which the response remains stable becomes wider as the modulation phase approaches $\pi$. 

As the number of modulated units increases, the number of UMFs increases and the unstable region nearly covers the entire range of modulation frequencies corresponding to $\beta=1$; {\it i.e.} $2\le\Omega_m<2\sqrt{1+4K_c}$. A semi-triangular region of stable response also forms in the frequency gap between UMFs with $\beta=1$ and $\beta=2$. Typically, the largest value of modulation amplitude at which the response remains stable occurs within this range of modulation frequencies. At lower modulation frequencies ($\beta>1$), several very small regions of stability appear sporadically within the unstable regions and the transition between unstable and stable regions becomes fragmented, narrow and dense. A main consistent trend within this parameter range is that the onset of instability occurs at higher modulation amplitudes as the modulation frequency decreases. 
 
Damping has an overall stabilizing effect by increasing the threshold of modulation amplitude that leads to parametric instability. This threshold is finite even at UMFs, in contrast to undamped systems. Damping has a stronger influence on stability diagram at lower modulation frequencies ($\beta>1$) and for longer systems. As a result, a damped system with several modulated units can remain stable at relatively high-amplitude modulations of low frequency. 

Parametric instability in systems with several degrees of freedom has not been explored extensively in the literature. Thus, there are several aspects of the phenomenon that warrant further detailed investigation. We found it infeasible to explore every detail within the confines of a single investigation. The intricate details of the stability diagrams at lower frequencies, the various effects of the modulation phase (wavenumber), and the influence of the number of units are not presented elsewhere in the literature, to the best of our knowledge. We hope that this initial exploration of parametric instability in spatiotemporally modulated systems encourages further analysis of the phenomenon and inspires future research on systems with high-amplitude modulations.

\begin{acknowledgments}
We are grateful to Andrus Giraldo, Korea Institute for Advanced Study, for helpful discussions on resonance tongues and for validating some of our preliminary stability diagrams using numerical continuation. We acknowledge financial support from the Natural Sciences and Engineering Research Council of Canada through the Discovery Grant program. J.W. acknowledges additional support from Concordia University and from Centre de Recherches Mathématique, Quebec.
\end{acknowledgments}

\appendix

\section{Non-dimensionalization} 
\label{appendix:nondimensionalization}
The equations of motion which govern the system of $n$ modulated units in Fig.~\ref{fig_nDoF} are:
\begin{equation}
\label{eqA1}
\begin{array}{c}
\displaystyle m \dv[2]{u_1}{t} + c \dv{u_1}{t} + k_1 u_1 
+ k_c \left(u_{1}-u_{2}\right) = 0,\\
\vdots \\
\displaystyle m \dv[2]{u_p}{t} + c \dv{u_p}{t} + k_p u_p 
+ k_c \left(2u_p-u_{p-1}-u_{p+1}\right) = 0,\\
\vdots \\
\displaystyle m \dv[2]{u_n}{t} + c \dv{u_n}{t} + k_n u_n 
+ k_c \left(u_{n}-u_{n-1}\right) = 0,
\end{array}
\end{equation}
where $k_p=k_{g,DC}+k_{g,AC}\cos(\omega_m t - \phi_p)$ and $\phi_p=(p-1)\phi$ for $p=1,2,\cdots,n$. 
We use $\tau=\omega_0 t$ as the nondimensional time with $\omega_0=\sqrt{k_{g,DC}/m}$. We define $\zeta=c/(2m\omega_0)$, $\Omega_m=\omega_m/\omega_0$, $\Omega_f=\omega_f/\omega_0$, $K_c=k_c/k_{g,DC}$ and $K_m=k_{g,AC}/k_{g,DC}$. The variable $x_p=u_p/a$ is used as nondimensional displacement where $a$ is a representative length.
After substituting these parameters into \eqN{eqA1}, we obtain:

\begin{align}
\label{eqA2}
m a \omega_0^2 \ddot{x}_p &+ ak_{g,DC} x_p \left[1 + K_m \cos{\left( \Omega_m \tau - \phi_p \right)} \right]\nonumber \\ 
&+2\zeta m a \omega_0^2 \dot{x}_p + a K_c k_{g,DC} \Delta^2_p = 0,
\end{align}
where $\ddot{x}_p$ and $\dot{x}_p$ represent $\dd[2]{x_p} / \dd[2]{\tau}$ and $\dd{x_p} / \dd{\tau}$ respectively. The difference terms are defined as $\Delta^2_1 = x_1 - x_2$ and $\Delta^2_n = x_n - x_{n-1}$ at two ends of the system, and $\Delta^2_p = 2 x_p - x_{p+1} - x_{p-1}$ elsewhere. Dividing the two sides of \eqN{eqA2} by $ak_{g,DC}$ yields \eqN{eq_EoM_p} in the main text. 
In this paper, calculations and analysis of the $n$-DoF modulated system are all based on the nondimensional form of the equations.

\section{Stability analysis using the harmonic balance method} 
\label{appendix:harmonic_balancing}

\begin{figure}[b!]
\centerline{\includegraphics[scale=0.45]{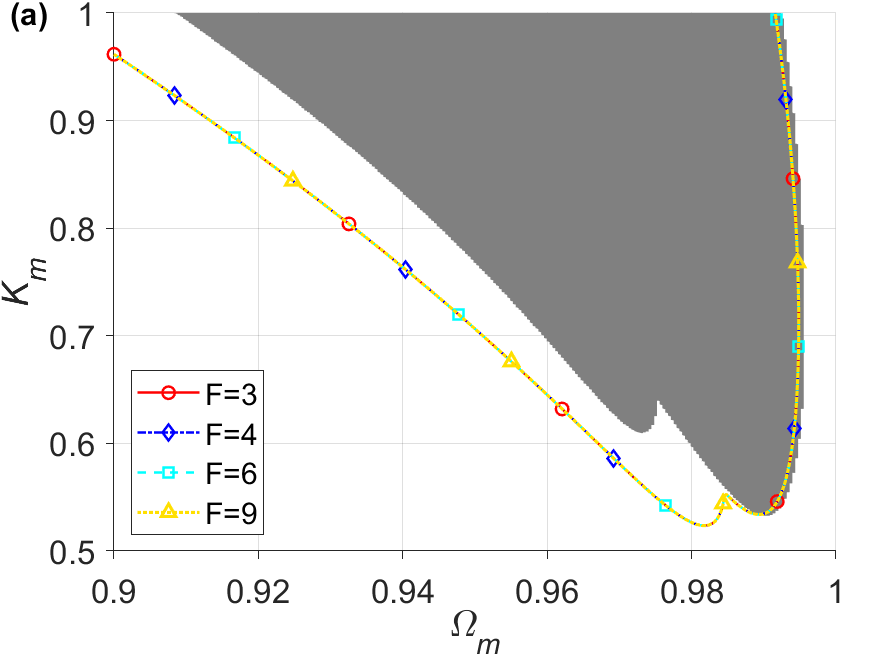}}
\centerline{\includegraphics[scale=0.45]{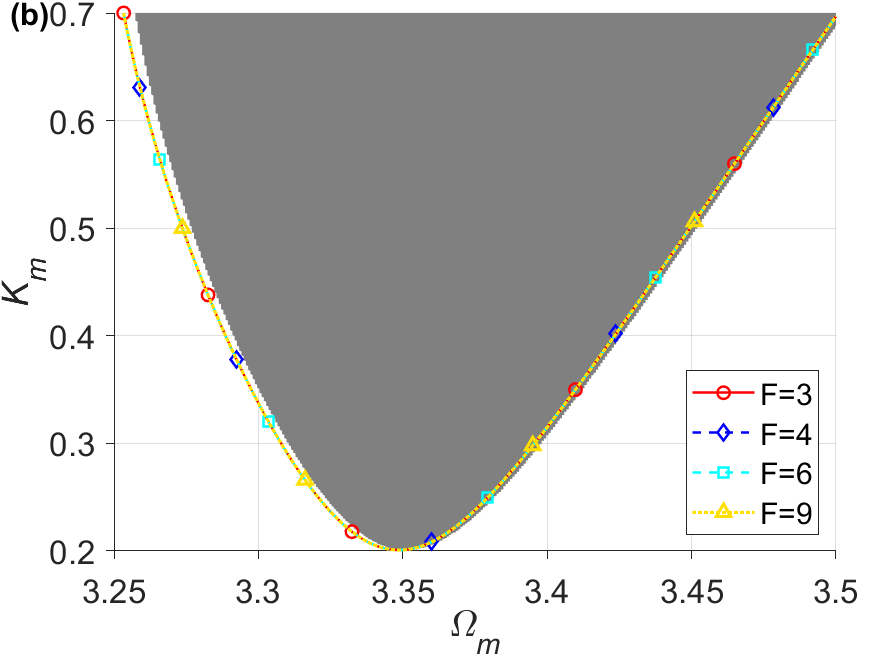}}
\caption{Transition curves calculated using the harmonic balance method. Grey regions indicate unstable response as determined by Floquet theory.
(a) $n=2$, $\zeta=0.01$, $\phi=0.5\pi$ and $K_c=0.6$; (b) $n=3$, $\zeta=0.02$, $\phi=0.5\pi$ and $K_c=0.6$.}
\label{fig_lambda}
\end{figure}

An approximated solution for Eq.~\eqref{eq_EoM_p} can be expressed as the product of a Fourier series and an exponential function:
\begin{equation}
\label{eq_HB1}
x_p(\tau) = \sum^{\infty}_{k=-\infty} [y_{p,k} e^{ik\Omega_m\tau}] e^{\lambda\tau}+c.c.
\end{equation}
where $p\in[1,n]$, $\lambda$ is the characteristic exponent and $c.c.$ represents the corresponding complex conjugate. 
Substitute Eq.~\eqref{eq_HB1} into Eq.~\eqref{eq_EoM_p} and balance the terms with harmonics $e^{(\lambda+ik\Omega_m)\tau}$, we obtain $n$ sets of equations in the form:
\begin{equation}
\label{eq_HB2}
\displaystyle \frac{K_m}{2}e^{-i\phi_p}y_{p,k-1}+c_k\,y_{p,k}+\frac{K_m}{2} e^{i\phi_p} y_{p,k+1}-K_c z_{p,k}=0
\end{equation}
where $c_k=(\lambda+ik\Omega_m)^2+2\zeta(\lambda+ik\Omega_m)+1+K_c$ for $p\in\{1,n\}$, $z_{1,k}=y_{2,k}$, $z_{n,k}=y_{n-1,k}$, and $c_k=(\lambda+ik\Omega_m)^2+2\zeta(\lambda+ik\Omega_m)+1+2K_c$ and $z_{p,k}=y_{p-1,k}+y_{p+1,k}$ for $p\in[2,n-1]$. Eq.~(\ref{eq_HB2}) can be written in matrix form as:
\begin{equation}
\label{eq_HB3}
\underline{\underline{M}}(\lambda,\phi,\zeta,K_m,K_c)\,\,\underline{Y}=\underline{O}
\end{equation}
where $\underline{O}$ is a zero vector and $\underline{Y}=[\cdots,y_{n,-1},y_{n,0},y_{n,1},\cdots]^T$  is the vector of unknown complex amplitudes. Matrix $\underline{\underline{M}}$ can be written as:
\begin{equation}
\underline{\underline{M}}=\begin{bmatrix}
\underline{\underline{M_1}} & \underline{\underline{K}} & \underline{\underline{O}} & \cdots & \underline{\underline{O}}\\
\underline{\underline{K}} & \underline{\underline{M_2}} & \underline{\underline{K}} & \cdots & \underline{\underline{O}}\\
\underline{\underline{O}} & \underline{\underline{K}} & \underline{\underline{M_3}} & \cdots & \underline{\underline{O}}\\
\vdots & \vdots & \vdots & \ddots & \vdots\\
\underline{\underline{O}} & \underline{\underline{O}} & \underline{\underline{O}} & \cdots & \underline{\underline{M_n}}
\end{bmatrix}
\end{equation}
where $\underline{\underline{O}}$ is a zero matrix and $\underline{\underline{K}}$ is a diagonal matrix in which all the elements in the main diagonal are $-K_c$. Matrix $\underline{\underline{M_p}}$ is:
\begin{equation}
\underline{\underline{M_p}}=\begin{bmatrix}
\ddots & \ddots & \ddots & \ddots & \ddots & & \\
\cdots & 0 & \frac{K_m}{2}e^{-i\phi_p} & c_k & \frac{K_m}{2}e^{i\phi_p} & 0 & \cdots\\
 & & \ddots & \ddots & \ddots & \ddots & \ddots
\end{bmatrix}
\end{equation}
If a non-zero solution for $\underline{Y}$ exists, then the determinant of $\underline{\underline{M}}$ must be zero. For each given combination of $\phi$, $\zeta$, $K_m$ and $K_c$, the unknown characteristic exponent $\lambda$ can be solved from $\det(\underline{\underline{M}})=0$. If the real part of every solution for $\det(\underline{\underline{M}}(\lambda))=0$ is negative or zero, then the response of the system in Eq.~\eqref{eq_EoM_p} is stable; otherwise, the response of the system is unstable.

The expansion in Eq.~\eqref{eq_HB1} assumes that every mode has the same characteristic exponent, $\lambda$. This would mean that the amplitude of every mode component grows exponentially at the same rate when instability occurs. However, this cannot be guaranteed. We validate the accuracy of this assumption by comparing its predicted stability charts to the results calculated numerically based on Floquet theory. 

In practice, the infinite summation in Eq.~\eqref{eq_HB1} is truncated at a finite value of $k$, for example $k\in[-F,F]$ with $F\in\mathbb{N}$. Fig.~\ref{fig_lambda} shows the transition curves calculated for a 2-DoF system and a 3-DoF system with $F\in\{3,4,6,9\}$. The grey areas indicate parameter values that lead to unstable results as determined by Floquet theory. In both examples, the value of $F$ has little effect on the transition curves calculated by the harmonic balance method for $F\geq3$. We have used the following identities to reduce the cost of calculation: $\det(\underline{\underline{M}})=\det(\underline{\underline{M_1}}\,\underline{\underline{M_2}}-\underline{\underline{K}}\,\underline{\underline{K}})$ for $n=2$ and $\det(\underline{\underline{M}})=\det(\underline{\underline{M_1}}\,\underline{\underline{M_2}}\,\underline{\underline{M_3}}-\underline{\underline{K}}\,\underline{\underline{K}}(\underline{\underline{M_1}}+\underline{\underline{M_3}}))$ for $n=3$. 

Fig.~\ref{fig_lambda} shows that the transition curves calculated based on the harmonic balance method do not match well with those computed using Floquet theory. The main shortcoming is in assuming the same exponent for different modes. It is possible to update the expansion in Eq.(\ref{eq_HB1}) to account for this. Instead, we use direct numerical computation of the onset of parametric instability based on Floquet theory, as explained in Section~\ref{subsec1}.

\section{Sets of harmonic terms at different frequencies} 
\label{appendix:harmonic_terms}

The sets of harmonic terms in~\eqN{eq_EoM_2DoF_e1_freqs} are:
\begin{widetext}
\begin{align*}
\mathcal{T}_1 \left[\frac{\Omega_{2,1}}{\Omega_m}\right] &= 2\frac{\Omega_{2,1}}{\Omega_m} \dv{A_1}{\nu} \sin{\frac{\Omega_{2,1}}{\Omega_m} \mu} - 2\frac{\Omega_{2,1}}{\Omega_m} \dv{B_1}{\nu} \cos{\frac{\Omega_{2,1}}{\Omega_m} \mu}, \\
\mathcal{T}_1 \left[\beta + \frac{\Omega_{2,1}}{\Omega_m}\right] &= \left[ -\frac{\left( 1+\cos{\phi} \right) A_1}{4\Omega_m^2} + \frac{\sin{\phi}B_1}{4\Omega_m^2} \right] \cos{\left(\beta + \frac{\Omega_{2,1}}{\Omega_m}\right)\mu} - \left[ \frac{\left( 1+\cos{\phi} \right) B_1}{4\Omega_m^2} + \frac{\sin{\phi}A_1}{4\Omega_m^2} \right] \sin{\left(\beta + \frac{\Omega_{2,1}}{\Omega_m}\right)\mu}, \\
\mathcal{T}_1 \left[\beta - \frac{\Omega_{2,1}}{\Omega_m}\right] &= -\left[ \frac{\left( 1+\cos{\phi} \right) A_1}{4\Omega_m^2} + \frac{\sin{\phi}B_1}{4\Omega_m^2} \right] \cos{\left(\beta - \frac{\Omega_{2,1}}{\Omega_m}\right)\mu} + \left[ \frac{\left( 1+\cos{\phi} \right) B_1}{4\Omega_m^2} - \frac{\sin{\phi}A_1}{4\Omega_m^2} \right] \sin{\left(\beta - \frac{\Omega_{2,1}}{\Omega_m}\right)\mu}, \\
\mathcal{T}_1 \left[\beta + \frac{\Omega_{2,2}}{\Omega_m}\right] &= -\left[ \frac{\left( 1-\cos{\phi} \right) A_2}{4\Omega_m^2} + \frac{\sin{\phi}B_2}{4\Omega_m^2} \right] \cos{\left(\beta + \frac{\Omega_{2,2}}{\Omega_m}\right)\mu} + \left[ -\frac{\left( 1-\cos{\phi} \right) B_2}{4\Omega_m^2} + \frac{\sin{\phi}A_2}{4\Omega_m^2} \right] \sin{\left(\beta + \frac{\Omega_{2,2}}{\Omega_m}\right)\mu}, \\
\mathcal{T}_1 \left[\beta - \frac{\Omega_{2,2}}{\Omega_m}\right] &= -\left[ \frac{\left( 1-\cos{\phi} \right) A_2}{4\Omega_m^2} - \frac{\sin{\phi}B_2}{4\Omega_m^2} \right] \cos{\left(\beta - \frac{\Omega_{2,2}}{\Omega_m}\right)\mu} + \left[ \frac{\left( 1-\cos{\phi} \right) B_2}{4\Omega_m^2} + \frac{\sin{\phi}A_2}{4\Omega_m^2} \right] \sin{\left(\beta - \frac{\Omega_{2,2}}{\Omega_m}\right)\mu},
\end{align*}
\end{widetext}
and 
\begin{widetext}
\begin{align*}
\mathcal{T}_2 \left[\frac{\Omega_{2,2}}{\Omega_m}\right] &= 2\frac{\Omega_{2,2}}{\Omega_m} \dv{A_2}{\nu} \sin{\frac{\Omega_{2,2}}{\Omega_m} \mu} - 2\frac{\Omega_{2,2}}{\Omega_m} \dv{B_2}{\nu} \cos{\frac{\Omega_{2,2}}{\Omega_m} \mu}, \\
\mathcal{T}_2 \left[\beta + \frac{\Omega_{2,1}}{\Omega_m}\right] &= \left[ \frac{\left( -1+\cos{\phi} \right) A_1}{4\Omega_m^2} - \frac{\sin{\phi}B_1}{4\Omega_m^2} \right] \cos{\left(\beta + \frac{\Omega_{2,1}}{\Omega_m}\right)\mu} + \left[ \frac{\left( -1+\cos{\phi} \right) B_1}{4\Omega_m^2} + \frac{\sin{\phi}A_1}{4\Omega_m^2} \right] \sin{\left(\beta + \frac{\Omega_{2,1}}{\Omega_m}\right)\mu}, \\
\mathcal{T}_2 \left[\beta - \frac{\Omega_{2,1}}{\Omega_m}\right] &= \left[ \frac{\left( -1+\cos{\phi} \right) A_1}{4\Omega_m^2} + \frac{\sin{\phi}B_1}{4\Omega_m^2} \right] \cos{\left(\beta - \frac{\Omega_{2,1}}{\Omega_m}\right)\mu} + \left[ -\frac{\left( -1+\cos{\phi} \right) B_1}{4\Omega_m^2} + \frac{\sin{\phi}A_1}{4\Omega_m^2} \right] \sin{\left(\beta - \frac{\Omega_{2,1}}{\Omega_m}\right)\mu}, \\
\mathcal{T}_2 \left[\beta + \frac{\Omega_{2,2}}{\Omega_m}\right] &= \left[ \frac{\left( -1-\cos{\phi} \right) A_2}{4\Omega_m^2} + \frac{\sin{\phi}B_2}{4\Omega_m^2} \right] \cos{\left(\beta + \frac{\Omega_{2,2}}{\Omega_m}\right)\mu} + \left[ \frac{\left( -1-\cos{\phi} \right) B_2}{4\Omega_m^2} - \frac{\sin{\phi}A_2}{4\Omega_m^2} \right] \sin{\left(\beta + \frac{\Omega_{2,2}}{\Omega_m}\right)\mu}, \\
\mathcal{T}_2 \left[\beta - \frac{\Omega_{2,2}}{\Omega_m}\right] &= \left[ \frac{\left( -1-\cos{\phi} \right) A_2}{4\Omega_m^2} - \frac{\sin{\phi}B_2}{4\Omega_m^2} \right] \cos{\left(\beta - \frac{\Omega_{2,2}}{\Omega_m}\right)\mu} + \left[ \frac{\left( 1+\cos{\phi} \right) B_2}{4\Omega_m^2} + \frac{\sin{\phi}A_2}{4\Omega_m^2} \right] \sin{\left(\beta - \frac{\Omega_{2,2}}{\Omega_m}\right)\mu},
\end{align*}
\end{widetext}


\bibliography{myrefs}

\providecommand{\noopsort}[1]{}\providecommand{\singleletter}[1]{#1}%
\begin{thebibliography}{43}%
\makeatletter
\providecommand \@ifxundefined [1]{%
 \@ifx{#1\undefined}
}%
\providecommand \@ifnum [1]{%
 \ifnum #1\expandafter \@firstoftwo
 \else \expandafter \@secondoftwo
 \fi
}%
\providecommand \@ifx [1]{%
 \ifx #1\expandafter \@firstoftwo
 \else \expandafter \@secondoftwo
 \fi
}%
\providecommand \natexlab [1]{#1}%
\providecommand \enquote  [1]{``#1''}%
\providecommand \bibnamefont  [1]{#1}%
\providecommand \bibfnamefont [1]{#1}%
\providecommand \citenamefont [1]{#1}%
\providecommand \href@noop [0]{\@secondoftwo}%
\providecommand \href [0]{\begingroup \@sanitize@url \@href}%
\providecommand \@href[1]{\@@startlink{#1}\@@href}%
\providecommand \@@href[1]{\endgroup#1\@@endlink}%
\providecommand \@sanitize@url [0]{\catcode `\\12\catcode `\$12\catcode `\&12\catcode `\#12\catcode `\^12\catcode `\_12\catcode `\%12\relax}%
\providecommand \@@startlink[1]{}%
\providecommand \@@endlink[0]{}%
\providecommand \url  [0]{\begingroup\@sanitize@url \@url }%
\providecommand \@url [1]{\endgroup\@href {#1}{\urlprefix }}%
\providecommand \urlprefix  [0]{URL }%
\providecommand \Eprint [0]{\href }%
\providecommand \doibase [0]{https://doi.org/}%
\providecommand \selectlanguage [0]{\@gobble}%
\providecommand \bibinfo  [0]{\@secondoftwo}%
\providecommand \bibfield  [0]{\@secondoftwo}%
\providecommand \translation [1]{[#1]}%
\providecommand \BibitemOpen [0]{}%
\providecommand \bibitemStop [0]{}%
\providecommand \bibitemNoStop [0]{.\EOS\space}%
\providecommand \EOS [0]{\spacefactor3000\relax}%
\providecommand \BibitemShut  [1]{\csname bibitem#1\endcsname}%
\let\auto@bib@innerbib\@empty
\bibitem [{\citenamefont {Lurie}(2017)}]{KLurie_2017}%
  \BibitemOpen
  \bibfield  {author} {\bibinfo {author} {\bibfnamefont {K.~A.}\ \bibnamefont {Lurie}},\ }\href@noop {} {\emph {\bibinfo {title} {An introduction to the mathematical theory of dynamic materials}}},\ Vol.~\bibinfo {volume} {15}\ (\bibinfo  {publisher} {Springer},\ \bibinfo {year} {2017})\BibitemShut {NoStop}%
\bibitem [{\citenamefont {Nassar}\ \emph {et~al.}(2020)\citenamefont {Nassar}, \citenamefont {Yousefzadeh}, \citenamefont {Fleury}, \citenamefont {Ruzzene}, \citenamefont {Alu}, \citenamefont {Daraio}, \citenamefont {Norris}, \citenamefont {Huang},\ and\ \citenamefont {Haberman}}]{HNassar_2020}%
  \BibitemOpen
  \bibfield  {author} {\bibinfo {author} {\bibfnamefont {H.}~\bibnamefont {Nassar}}, \bibinfo {author} {\bibfnamefont {B.}~\bibnamefont {Yousefzadeh}}, \bibinfo {author} {\bibfnamefont {R.}~\bibnamefont {Fleury}}, \bibinfo {author} {\bibfnamefont {M.}~\bibnamefont {Ruzzene}}, \bibinfo {author} {\bibfnamefont {A.}~\bibnamefont {Alu}}, \bibinfo {author} {\bibfnamefont {C.}~\bibnamefont {Daraio}}, \bibinfo {author} {\bibfnamefont {A.}~\bibnamefont {Norris}}, \bibinfo {author} {\bibfnamefont {G.}~\bibnamefont {Huang}},\ and\ \bibinfo {author} {\bibfnamefont {M.}~\bibnamefont {Haberman}},\ }\bibfield  {title} {\bibinfo {title} {Nonreciprocity in acoustic and elastic materials},\ }\href@noop {} {\bibfield  {journal} {\bibinfo  {journal} {Nature Reviews Materials}\ }\textbf {\bibinfo {volume} {5}},\ \bibinfo {pages} {667–685} (\bibinfo {year} {2020})}\BibitemShut {NoStop}%
\bibitem [{\citenamefont {Nassar}\ \emph {et~al.}(2018)\citenamefont {Nassar}, \citenamefont {Chen}, \citenamefont {Norris},\ and\ \citenamefont {Huang}}]{HNassar_2018}%
  \BibitemOpen
  \bibfield  {author} {\bibinfo {author} {\bibfnamefont {H.}~\bibnamefont {Nassar}}, \bibinfo {author} {\bibfnamefont {H.}~\bibnamefont {Chen}}, \bibinfo {author} {\bibfnamefont {A.}~\bibnamefont {Norris}},\ and\ \bibinfo {author} {\bibfnamefont {G.}~\bibnamefont {Huang}},\ }\bibfield  {title} {\bibinfo {title} {Quantization of band tilting in modulated phononic crystals},\ }\href@noop {} {\bibfield  {journal} {\bibinfo  {journal} {Physical Review B}\ }\textbf {\bibinfo {volume} {97}},\ \bibinfo {pages} {014305} (\bibinfo {year} {2018})}\BibitemShut {NoStop}%
\bibitem [{\citenamefont {Hasan}\ \emph {et~al.}(2019)\citenamefont {Hasan}, \citenamefont {Calderin}, \citenamefont {Lucas}, \citenamefont {Runge},\ and\ \citenamefont {Deymier}}]{deymier_2019}%
  \BibitemOpen
  \bibfield  {author} {\bibinfo {author} {\bibfnamefont {M.~A.}\ \bibnamefont {Hasan}}, \bibinfo {author} {\bibfnamefont {L.}~\bibnamefont {Calderin}}, \bibinfo {author} {\bibfnamefont {P.}~\bibnamefont {Lucas}}, \bibinfo {author} {\bibfnamefont {K.}~\bibnamefont {Runge}},\ and\ \bibinfo {author} {\bibfnamefont {P.~A.}\ \bibnamefont {Deymier}},\ }\bibfield  {title} {\bibinfo {title} {Geometric phase invariance in spatiotemporal modulated elastic system},\ }\href@noop {} {\bibfield  {journal} {\bibinfo  {journal} {Journal of Sound and Vibration}\ }\textbf {\bibinfo {volume} {459}},\ \bibinfo {pages} {114843} (\bibinfo {year} {2019})}\BibitemShut {NoStop}%
\bibitem [{\citenamefont {Trainiti}\ and\ \citenamefont {Ruzzene}(2016)}]{NJP2016}%
  \BibitemOpen
  \bibfield  {author} {\bibinfo {author} {\bibfnamefont {G.}~\bibnamefont {Trainiti}}\ and\ \bibinfo {author} {\bibfnamefont {M.}~\bibnamefont {Ruzzene}},\ }\bibfield  {title} {\bibinfo {title} {Non-reciprocal elastic wave propagation in spatiotemporal periodic structures},\ }\href@noop {} {\bibfield  {journal} {\bibinfo  {journal} {New Journal of Physics}\ }\textbf {\bibinfo {volume} {18}},\ \bibinfo {pages} {083047} (\bibinfo {year} {2016})}\BibitemShut {NoStop}%
\bibitem [{\citenamefont {Moghaddaszadeh}\ \emph {et~al.}(2022)\citenamefont {Moghaddaszadeh}, \citenamefont {Attarzadeh}, \citenamefont {Aref},\ and\ \citenamefont {Nouh}}]{PRA2022_Nouh}%
  \BibitemOpen
  \bibfield  {author} {\bibinfo {author} {\bibfnamefont {M.}~\bibnamefont {Moghaddaszadeh}}, \bibinfo {author} {\bibfnamefont {M.}~\bibnamefont {Attarzadeh}}, \bibinfo {author} {\bibfnamefont {A.}~\bibnamefont {Aref}},\ and\ \bibinfo {author} {\bibfnamefont {M.}~\bibnamefont {Nouh}},\ }\bibfield  {title} {\bibinfo {title} {Complex spatiotemporal modulations and non-hermitian degeneracies in pt-symmetric phononic materials},\ }\href@noop {} {\bibfield  {journal} {\bibinfo  {journal} {Physical Review Applied}\ }\textbf {\bibinfo {volume} {18}},\ \bibinfo {pages} {044013} (\bibinfo {year} {2022})}\BibitemShut {NoStop}%
\bibitem [{\citenamefont {Wu}\ and\ \citenamefont {Yousefzadeh}(2025)}]{paper1}%
  \BibitemOpen
  \bibfield  {author} {\bibinfo {author} {\bibfnamefont {J.}~\bibnamefont {Wu}}\ and\ \bibinfo {author} {\bibfnamefont {B.}~\bibnamefont {Yousefzadeh}},\ }\bibfield  {title} {\bibinfo {title} {Linear nonreciprocal dynamics of coupled modulated systems},\ }\href@noop {} {\bibfield  {journal} {\bibinfo  {journal} {The Journal of the Acoustical Society of America}\ }\textbf {\bibinfo {volume} {157}},\ \bibinfo {pages} {1356–1367} (\bibinfo {year} {2025})}\BibitemShut {NoStop}%
\bibitem [{\citenamefont {Champneys}(2011)}]{Champneys}%
  \BibitemOpen
  \bibfield  {author} {\bibinfo {author} {\bibfnamefont {A.}~\bibnamefont {Champneys}},\ }\bibfield  {title} {\bibinfo {title} {Dynamics of parametric excitation},\ }in\ \href@noop {} {\emph {\bibinfo {booktitle} {Mathematics of Complexity and Dynamical Systems}}},\ \bibinfo {editor} {edited by\ \bibinfo {editor} {\bibfnamefont {R.~A.}\ \bibnamefont {Meyers}}}\ (\bibinfo  {publisher} {Springer},\ \bibinfo {address} {New York},\ \bibinfo {year} {2011})\ p.\ \bibinfo {pages} {183–204}\BibitemShut {NoStop}%
\bibitem [{\citenamefont {Kovacic}\ \emph {et~al.}(2018)\citenamefont {Kovacic}, \citenamefont {Rand},\ and\ \citenamefont {Sah}}]{IKovacic_2018}%
  \BibitemOpen
  \bibfield  {author} {\bibinfo {author} {\bibfnamefont {I.}~\bibnamefont {Kovacic}}, \bibinfo {author} {\bibfnamefont {R.}~\bibnamefont {Rand}},\ and\ \bibinfo {author} {\bibfnamefont {S.~M.}\ \bibnamefont {Sah}},\ }\bibfield  {title} {\bibinfo {title} {Mathieu's equation and its generalizations: Overview of stability charts and their features},\ }\href@noop {} {\bibfield  {journal} {\bibinfo  {journal} {Applied Mechanics Reviews}\ }\textbf {\bibinfo {volume} {70}},\ \bibinfo {pages} {020802} (\bibinfo {year} {2018})}\BibitemShut {NoStop}%
\bibitem [{\citenamefont {Trainiti}\ \emph {et~al.}(2019)\citenamefont {Trainiti}, \citenamefont {Xia}, \citenamefont {Marconi}, \citenamefont {Cazzulani}, \citenamefont {Erturk},\ and\ \citenamefont {Ruzzene}}]{GTrainiti_2019}%
  \BibitemOpen
  \bibfield  {author} {\bibinfo {author} {\bibfnamefont {G.}~\bibnamefont {Trainiti}}, \bibinfo {author} {\bibfnamefont {Y.}~\bibnamefont {Xia}}, \bibinfo {author} {\bibfnamefont {J.}~\bibnamefont {Marconi}}, \bibinfo {author} {\bibfnamefont {G.}~\bibnamefont {Cazzulani}}, \bibinfo {author} {\bibfnamefont {A.}~\bibnamefont {Erturk}},\ and\ \bibinfo {author} {\bibfnamefont {M.}~\bibnamefont {Ruzzene}},\ }\bibfield  {title} {\bibinfo {title} {Time-periodic stiffness modulation in elastic metamaterials for selective wave filtering: Theory and experiment},\ }\href@noop {} {\bibfield  {journal} {\bibinfo  {journal} {Physical review letters}\ }\textbf {\bibinfo {volume} {122}},\ \bibinfo {pages} {124301} (\bibinfo {year} {2019})}\BibitemShut {NoStop}%
\bibitem [{\citenamefont {Thomes}\ \emph {et~al.}(2023)\citenamefont {Thomes}, \citenamefont {Beli}, \citenamefont {Sugino}, \citenamefont {Erturk},\ and\ \citenamefont {Junior}}]{RThomes_2023}%
  \BibitemOpen
  \bibfield  {author} {\bibinfo {author} {\bibfnamefont {R.~L.}\ \bibnamefont {Thomes}}, \bibinfo {author} {\bibfnamefont {D.}~\bibnamefont {Beli}}, \bibinfo {author} {\bibfnamefont {C.}~\bibnamefont {Sugino}}, \bibinfo {author} {\bibfnamefont {A.}~\bibnamefont {Erturk}},\ and\ \bibinfo {author} {\bibfnamefont {C.~D.~M.}\ \bibnamefont {Junior}},\ }\bibfield  {title} {\bibinfo {title} {Programmable moving defect for spatiotemporal wave localization in piezoelectric metamaterials},\ }\href@noop {} {\bibfield  {journal} {\bibinfo  {journal} {Physical Review Applied}\ }\textbf {\bibinfo {volume} {19}},\ \bibinfo {pages} {064031} (\bibinfo {year} {2023})}\BibitemShut {NoStop}%
\bibitem [{\citenamefont {Wang}\ \emph {et~al.}(2018)\citenamefont {Wang}, \citenamefont {Yousefzadeh}, \citenamefont {Chen}, \citenamefont {Nassar}, \citenamefont {Huang},\ and\ \citenamefont {Daraio}}]{YWang_2018}%
  \BibitemOpen
  \bibfield  {author} {\bibinfo {author} {\bibfnamefont {Y.}~\bibnamefont {Wang}}, \bibinfo {author} {\bibfnamefont {B.}~\bibnamefont {Yousefzadeh}}, \bibinfo {author} {\bibfnamefont {H.}~\bibnamefont {Chen}}, \bibinfo {author} {\bibfnamefont {H.}~\bibnamefont {Nassar}}, \bibinfo {author} {\bibfnamefont {G.}~\bibnamefont {Huang}},\ and\ \bibinfo {author} {\bibfnamefont {C.}~\bibnamefont {Daraio}},\ }\bibfield  {title} {\bibinfo {title} {Observation of nonreciprocal wave propagation in a dynamic phononic lattice},\ }\href@noop {} {\bibfield  {journal} {\bibinfo  {journal} {Physical Review Letters}\ }\textbf {\bibinfo {volume} {121}},\ \bibinfo {pages} {194301} (\bibinfo {year} {2018})}\BibitemShut {NoStop}%
\bibitem [{\citenamefont {Wan}\ \emph {et~al.}(2022)\citenamefont {Wan}, \citenamefont {Cao}, \citenamefont {Zeng}, \citenamefont {Guo}, \citenamefont {Oudich},\ and\ \citenamefont {Assouar}}]{SWan_2022}%
  \BibitemOpen
  \bibfield  {author} {\bibinfo {author} {\bibfnamefont {S.}~\bibnamefont {Wan}}, \bibinfo {author} {\bibfnamefont {L.}~\bibnamefont {Cao}}, \bibinfo {author} {\bibfnamefont {Y.}~\bibnamefont {Zeng}}, \bibinfo {author} {\bibfnamefont {T.}~\bibnamefont {Guo}}, \bibinfo {author} {\bibfnamefont {M.}~\bibnamefont {Oudich}},\ and\ \bibinfo {author} {\bibfnamefont {B.}~\bibnamefont {Assouar}},\ }\bibfield  {title} {\bibinfo {title} {Low-frequency nonreciprocal flexural wave propagation via compact cascaded time-modulated resonators},\ }\href@noop {} {\bibfield  {journal} {\bibinfo  {journal} {Applied Physics Letters}\ }\textbf {\bibinfo {volume} {120}},\ \bibinfo {pages} {231701} (\bibinfo {year} {2022})}\BibitemShut {NoStop}%
\bibitem [{\citenamefont {Nayfeh}\ and\ \citenamefont {Mook}(1979)}]{ANayfeh_1979}%
  \BibitemOpen
  \bibfield  {author} {\bibinfo {author} {\bibfnamefont {A.~H.}\ \bibnamefont {Nayfeh}}\ and\ \bibinfo {author} {\bibfnamefont {D.~T.}\ \bibnamefont {Mook}},\ }\href@noop {} {\emph {\bibinfo {title} {Nonlinear oscillations}}}\ (\bibinfo  {publisher} {John Wiley \& Sons, Inc.},\ \bibinfo {year} {1979})\BibitemShut {NoStop}%
\bibitem [{\citenamefont {Curry}(1976)}]{swing}%
  \BibitemOpen
  \bibfield  {author} {\bibinfo {author} {\bibfnamefont {S.~M.}\ \bibnamefont {Curry}},\ }\bibfield  {title} {\bibinfo {title} {How children swing},\ }\href@noop {} {\bibfield  {journal} {\bibinfo  {journal} {American Journal of Physics}\ }\textbf {\bibinfo {volume} {44}},\ \bibinfo {pages} {924–926} (\bibinfo {year} {1976})}\BibitemShut {NoStop}%
\bibitem [{\citenamefont {Brimacombe}\ \emph {et~al.}(2021)\citenamefont {Brimacombe}, \citenamefont {Corless},\ and\ \citenamefont {Zamir}}]{MathieuEq_hist}%
  \BibitemOpen
  \bibfield  {author} {\bibinfo {author} {\bibfnamefont {C.}~\bibnamefont {Brimacombe}}, \bibinfo {author} {\bibfnamefont {R.~M.}\ \bibnamefont {Corless}},\ and\ \bibinfo {author} {\bibfnamefont {M.}~\bibnamefont {Zamir}},\ }\bibfield  {title} {\bibinfo {title} {Computation and applications of mathieu functions: A historical perspective},\ }\href@noop {} {\bibfield  {journal} {\bibinfo  {journal} {SIAM Review}\ }\textbf {\bibinfo {volume} {63}},\ \bibinfo {pages} {653–720} (\bibinfo {year} {2021})}\BibitemShut {NoStop}%
\bibitem [{\citenamefont {Mathieu}(1868)}]{Mathieu_1868}%
  \BibitemOpen
  \bibfield  {author} {\bibinfo {author} {\bibfnamefont {{\'E}.}~\bibnamefont {Mathieu}},\ }\bibfield  {title} {\bibinfo {title} {M{\'e}moire sur le mouvement vibratoire d'une membrane de forme elliptique},\ }\href@noop {} {\bibfield  {journal} {\bibinfo  {journal} {Journal de math{\'e}matiques pures et appliqu{\'e}es}\ }\textbf {\bibinfo {volume} {13}},\ \bibinfo {pages} {137–203} (\bibinfo {year} {1868})}\BibitemShut {NoStop}%
\bibitem [{\citenamefont {Rayleigh}(1883)}]{JRayleigh_1883}%
  \BibitemOpen
  \bibfield  {author} {\bibinfo {author} {\bibfnamefont {J.~W.~S.}\ \bibnamefont {Rayleigh}},\ }\bibfield  {title} {\bibinfo {title} {On maintained vibrations},\ }\href@noop {} {\bibfield  {journal} {\bibinfo  {journal} {The London, Edinburgh, and Dublin Philosophical Magazine and Journal of Science}\ }\textbf {\bibinfo {volume} {15}},\ \bibinfo {pages} {229–235} (\bibinfo {year} {1883})}\BibitemShut {NoStop}%
\bibitem [{\citenamefont {Rhoads}\ \emph {et~al.}(2008)\citenamefont {Rhoads}, \citenamefont {Miller}, \citenamefont {Shaw},\ and\ \citenamefont {Feeny}}]{JRhoads_2008}%
  \BibitemOpen
  \bibfield  {author} {\bibinfo {author} {\bibfnamefont {J.~F.}\ \bibnamefont {Rhoads}}, \bibinfo {author} {\bibfnamefont {N.~J.}\ \bibnamefont {Miller}}, \bibinfo {author} {\bibfnamefont {S.~W.}\ \bibnamefont {Shaw}},\ and\ \bibinfo {author} {\bibfnamefont {B.~F.}\ \bibnamefont {Feeny}},\ }\bibfield  {title} {\bibinfo {title} {Mechanical domain parametric amplification},\ }\href@noop {} {\bibfield  {journal} {\bibinfo  {journal} {Journal of Vibration and Acoustics}\ }\textbf {\bibinfo {volume} {130}},\ \bibinfo {pages} {061006} (\bibinfo {year} {2008})}\BibitemShut {NoStop}%
\bibitem [{\citenamefont {Moran}\ \emph {et~al.}(2013)\citenamefont {Moran}, \citenamefont {Burgner}, \citenamefont {Shaw},\ and\ \citenamefont {Turner}}]{SShaw_2018}%
  \BibitemOpen
  \bibfield  {author} {\bibinfo {author} {\bibfnamefont {K.}~\bibnamefont {Moran}}, \bibinfo {author} {\bibfnamefont {C.}~\bibnamefont {Burgner}}, \bibinfo {author} {\bibfnamefont {S.}~\bibnamefont {Shaw}},\ and\ \bibinfo {author} {\bibfnamefont {K.}~\bibnamefont {Turner}},\ }\bibfield  {title} {\bibinfo {title} {A review of parametric resonance in microelectromechanical systems},\ }\href@noop {} {\bibfield  {journal} {\bibinfo  {journal} {Nonlinear theory and its applications, IEICE}\ }\textbf {\bibinfo {volume} {4}},\ \bibinfo {pages} {198–224} (\bibinfo {year} {2013})}\BibitemShut {NoStop}%
\bibitem [{\citenamefont {De~felice}\ and\ \citenamefont {Sorrentino}(2022)}]{aero1}%
  \BibitemOpen
  \bibfield  {author} {\bibinfo {author} {\bibfnamefont {A.}~\bibnamefont {De~felice}}\ and\ \bibinfo {author} {\bibfnamefont {S.}~\bibnamefont {Sorrentino}},\ }\bibfield  {title} {\bibinfo {title} {Effects of anisotropic supports on the stability of parametrically excited slender rotors},\ }\href@noop {} {\bibfield  {journal} {\bibinfo  {journal} {Nonlinear Dynamics}\ }\textbf {\bibinfo {volume} {109}},\ \bibinfo {pages} {793–813} (\bibinfo {year} {2022})}\BibitemShut {NoStop}%
\bibitem [{\citenamefont {De~Felice}\ and\ \citenamefont {Sorrentino}(2021)}]{aero2}%
  \BibitemOpen
  \bibfield  {author} {\bibinfo {author} {\bibfnamefont {A.}~\bibnamefont {De~Felice}}\ and\ \bibinfo {author} {\bibfnamefont {S.}~\bibnamefont {Sorrentino}},\ }\bibfield  {title} {\bibinfo {title} {Damping and gyroscopic effects on the stability of parametrically excited continuous rotor systems},\ }\href@noop {} {\bibfield  {journal} {\bibinfo  {journal} {Nonlinear Dynamics}\ }\textbf {\bibinfo {volume} {103}},\ \bibinfo {pages} {3529–3555} (\bibinfo {year} {2021})}\BibitemShut {NoStop}%
\bibitem [{\citenamefont {Acar}\ \emph {et~al.}(2020)\citenamefont {Acar}, \citenamefont {Acar},\ and\ \citenamefont {Feeny}}]{BFeeny_2020}%
  \BibitemOpen
  \bibfield  {author} {\bibinfo {author} {\bibfnamefont {G.~D.}\ \bibnamefont {Acar}}, \bibinfo {author} {\bibfnamefont {M.~A.}\ \bibnamefont {Acar}},\ and\ \bibinfo {author} {\bibfnamefont {B.~F.}\ \bibnamefont {Feeny}},\ }\bibfield  {title} {\bibinfo {title} {Parametric resonances of a three-blade-rotor system with reference to wind turbines},\ }\href@noop {} {\bibfield  {journal} {\bibinfo  {journal} {Journal of Vibration and Acoustics}\ }\textbf {\bibinfo {volume} {142}},\ \bibinfo {pages} {021013} (\bibinfo {year} {2020})}\BibitemShut {NoStop}%
\bibitem [{\citenamefont {Yao}\ \emph {et~al.}(2011)\citenamefont {Yao}, \citenamefont {Mei},\ and\ \citenamefont {Chen}}]{chatter1}%
  \BibitemOpen
  \bibfield  {author} {\bibinfo {author} {\bibfnamefont {Z.}~\bibnamefont {Yao}}, \bibinfo {author} {\bibfnamefont {D.}~\bibnamefont {Mei}},\ and\ \bibinfo {author} {\bibfnamefont {Z.}~\bibnamefont {Chen}},\ }\bibfield  {title} {\bibinfo {title} {Chatter suppression by parametric excitation: Model and experiments},\ }\href@noop {} {\bibfield  {journal} {\bibinfo  {journal} {Journal of Sound and Vibration}\ }\textbf {\bibinfo {volume} {330}},\ \bibinfo {pages} {2995–3005} (\bibinfo {year} {2011})}\BibitemShut {NoStop}%
\bibitem [{\citenamefont {Turkes}\ \emph {et~al.}(2017)\citenamefont {Turkes}, \citenamefont {Orak}, \citenamefont {Neeli}, \citenamefont {Sahin},\ and\ \citenamefont {Selvi}}]{chatter2}%
  \BibitemOpen
  \bibfield  {author} {\bibinfo {author} {\bibfnamefont {E.}~\bibnamefont {Turkes}}, \bibinfo {author} {\bibfnamefont {S.}~\bibnamefont {Orak}}, \bibinfo {author} {\bibfnamefont {S.}~\bibnamefont {Neeli}}, \bibinfo {author} {\bibfnamefont {M.}~\bibnamefont {Sahin}},\ and\ \bibinfo {author} {\bibfnamefont {S.}~\bibnamefont {Selvi}},\ }\bibfield  {title} {\bibinfo {title} {Modelling of dynamic cutting force coefficients and chatter stability dependent on shear angle oscillation},\ }\href@noop {} {\bibfield  {journal} {\bibinfo  {journal} {International Journal of Advanced Manufacturing Technology}\ }\textbf {\bibinfo {volume} {91}},\ \bibinfo {pages} {679–686} (\bibinfo {year} {2017})}\BibitemShut {NoStop}%
\bibitem [{\citenamefont {St{\'e}p{\'a}n}\ and\ \citenamefont {Insperger}(2006)}]{control1}%
  \BibitemOpen
  \bibfield  {author} {\bibinfo {author} {\bibfnamefont {G.}~\bibnamefont {St{\'e}p{\'a}n}}\ and\ \bibinfo {author} {\bibfnamefont {T.}~\bibnamefont {Insperger}},\ }\bibfield  {title} {\bibinfo {title} {Stability of time-periodic and delayed systems - a route to act-and-wait control},\ }\href@noop {} {\bibfield  {journal} {\bibinfo  {journal} {Annual Reviews in Control}\ }\textbf {\bibinfo {volume} {30}},\ \bibinfo {pages} {159–168} (\bibinfo {year} {2006})}\BibitemShut {NoStop}%
\bibitem [{\citenamefont {Nassar}\ \emph {et~al.}(2024)\citenamefont {Nassar}, \citenamefont {Norris},\ and\ \citenamefont {Huang}}]{nassar_python}%
  \BibitemOpen
  \bibfield  {author} {\bibinfo {author} {\bibfnamefont {H.}~\bibnamefont {Nassar}}, \bibinfo {author} {\bibfnamefont {A.~N.}\ \bibnamefont {Norris}},\ and\ \bibinfo {author} {\bibfnamefont {G.}~\bibnamefont {Huang}},\ }\bibfield  {title} {\bibinfo {title} {Waves over a periodic progressive modulation: A python tutorial},\ }in\ \href@noop {} {\emph {\bibinfo {booktitle} {Acoustic Metamaterials: Absorption, Cloaking, Imaging, Time-Modulated Media, and Topological Crystals}}}\ (\bibinfo  {publisher} {Springer},\ \bibinfo {year} {2024})\ p.\ \bibinfo {pages} {505–533}\BibitemShut {NoStop}%
\bibitem [{\citenamefont {Jin}\ \emph {et~al.}(2024)\citenamefont {Jin}, \citenamefont {Li}, \citenamefont {Djafari-Rouhani}, \citenamefont {Torrent}, \citenamefont {Xiang},\ and\ \citenamefont {Xuan}}]{YJin_2024}%
  \BibitemOpen
  \bibfield  {author} {\bibinfo {author} {\bibfnamefont {Y.}~\bibnamefont {Jin}}, \bibinfo {author} {\bibfnamefont {W.}~\bibnamefont {Li}}, \bibinfo {author} {\bibfnamefont {B.}~\bibnamefont {Djafari-Rouhani}}, \bibinfo {author} {\bibfnamefont {D.}~\bibnamefont {Torrent}}, \bibinfo {author} {\bibfnamefont {Y.}~\bibnamefont {Xiang}},\ and\ \bibinfo {author} {\bibfnamefont {F.-Z.}\ \bibnamefont {Xuan}},\ }\bibfield  {title} {\bibinfo {title} {Exceptional points in time-varying oscillators with enhanced sensing sensitivity},\ }\href@noop {} {\bibfield  {journal} {\bibinfo  {journal} {Physical Review Applied}\ }\textbf {\bibinfo {volume} {22}},\ \bibinfo {pages} {034026} (\bibinfo {year} {2024})}\BibitemShut {NoStop}%
\bibitem [{\citenamefont {Cesari}(1963)}]{LCesari_1963}%
  \BibitemOpen
  \bibfield  {author} {\bibinfo {author} {\bibfnamefont {L.}~\bibnamefont {Cesari}},\ }\href@noop {} {\emph {\bibinfo {title} {Asymptotic behavior and stability problems in ordinary differential equations}}},\ \bibinfo {edition} {2nd}\ ed.\ (\bibinfo  {publisher} {Academic Press Inc.},\ \bibinfo {year} {1963})\BibitemShut {NoStop}%
\bibitem [{\citenamefont {Hsu}(1963)}]{CHsu_1963}%
  \BibitemOpen
  \bibfield  {author} {\bibinfo {author} {\bibfnamefont {C.~S.}\ \bibnamefont {Hsu}},\ }\bibfield  {title} {\bibinfo {title} {On the parametric excitation of a dynamic system having multiple degrees of freedom},\ }\href@noop {} {\bibfield  {journal} {\bibinfo  {journal} {Journal of Applied Mechanics}\ }\textbf {\bibinfo {volume} {30}},\ \bibinfo {pages} {367–372} (\bibinfo {year} {1963})}\BibitemShut {NoStop}%
\bibitem [{\citenamefont {Ince}(1927)}]{Ince_1927}%
  \BibitemOpen
  \bibfield  {author} {\bibinfo {author} {\bibfnamefont {E.~L.}\ \bibnamefont {Ince}},\ }\bibfield  {title} {\bibinfo {title} {Researches into the characteristic numbers of the mathieu equation},\ }\href@noop {} {\bibfield  {journal} {\bibinfo  {journal} {Proceedings of the Royal Society of Edinburgh}\ }\textbf {\bibinfo {volume} {46}},\ \bibinfo {pages} {20–29} (\bibinfo {year} {1927})}\BibitemShut {NoStop}%
\bibitem [{\citenamefont {Barakat}\ \emph {et~al.}(2023)\citenamefont {Barakat}, \citenamefont {Weig},\ and\ \citenamefont {Hagedorn}}]{Unstb_freq2}%
  \BibitemOpen
  \bibfield  {author} {\bibinfo {author} {\bibfnamefont {A.~A.}\ \bibnamefont {Barakat}}, \bibinfo {author} {\bibfnamefont {E.~M.}\ \bibnamefont {Weig}},\ and\ \bibinfo {author} {\bibfnamefont {P.}~\bibnamefont {Hagedorn}},\ }\bibfield  {title} {\bibinfo {title} {Non-trivial solutions and their stability in a two-degree-of-freedom mathieu–duffing system},\ }\href@noop {} {\bibfield  {journal} {\bibinfo  {journal} {Nonlinear Dynamics}\ }\textbf {\bibinfo {volume} {111}},\ \bibinfo {pages} {22119–22136} (\bibinfo {year} {2023})}\BibitemShut {NoStop}%
\bibitem [{\citenamefont {Deng}(2023)}]{Unstb_freq3}%
  \BibitemOpen
  \bibfield  {author} {\bibinfo {author} {\bibfnamefont {J.}~\bibnamefont {Deng}},\ }\bibfield  {title} {\bibinfo {title} {Numerical simulation of stability and responses of dynamic systems under parametric excitation},\ }\href@noop {} {\bibfield  {journal} {\bibinfo  {journal} {Applied Mathematical Modelling}\ }\textbf {\bibinfo {volume} {119}},\ \bibinfo {pages} {648–676} (\bibinfo {year} {2023})}\BibitemShut {NoStop}%
\bibitem [{\citenamefont {Ramirez-Barrios}\ \emph {et~al.}(2024)\citenamefont {Ramirez-Barrios}, \citenamefont {Collado},\ and\ \citenamefont {Dohnal}}]{Unstb_freq1}%
  \BibitemOpen
  \bibfield  {author} {\bibinfo {author} {\bibfnamefont {M.}~\bibnamefont {Ramirez-Barrios}}, \bibinfo {author} {\bibfnamefont {J.}~\bibnamefont {Collado}},\ and\ \bibinfo {author} {\bibfnamefont {F.}~\bibnamefont {Dohnal}},\ }\bibfield  {title} {\bibinfo {title} {Stability of periodic hamiltonian systems with equal dissipation},\ }\href@noop {} {\bibfield  {journal} {\bibinfo  {journal} {Nonlinear Dynamics}\ }\textbf {\bibinfo {volume} {112}},\ \bibinfo {pages} {17033–17053} (\bibinfo {year} {2024})}\BibitemShut {NoStop}%
\bibitem [{\citenamefont {Kim}\ \emph {et~al.}(2023)\citenamefont {Kim}, \citenamefont {Chong}, \citenamefont {Hajarolasvadi}, \citenamefont {Wang},\ and\ \citenamefont {Daraio}}]{BKim_2023}%
  \BibitemOpen
  \bibfield  {author} {\bibinfo {author} {\bibfnamefont {B.~L.}\ \bibnamefont {Kim}}, \bibinfo {author} {\bibfnamefont {C.}~\bibnamefont {Chong}}, \bibinfo {author} {\bibfnamefont {S.}~\bibnamefont {Hajarolasvadi}}, \bibinfo {author} {\bibfnamefont {Y.}~\bibnamefont {Wang}},\ and\ \bibinfo {author} {\bibfnamefont {C.}~\bibnamefont {Daraio}},\ }\bibfield  {title} {\bibinfo {title} {Dynamics of time-modulated, nonlinear phononic lattices},\ }\href@noop {} {\bibfield  {journal} {\bibinfo  {journal} {Physical Review E}\ }\textbf {\bibinfo {volume} {107}},\ \bibinfo {pages} {034211} (\bibinfo {year} {2023})}\BibitemShut {NoStop}%
\bibitem [{\citenamefont {Chong}\ \emph {et~al.}(2024)\citenamefont {Chong}, \citenamefont {Kim}, \citenamefont {Wallace},\ and\ \citenamefont {Daraio}}]{CChong_2024}%
  \BibitemOpen
  \bibfield  {author} {\bibinfo {author} {\bibfnamefont {C.}~\bibnamefont {Chong}}, \bibinfo {author} {\bibfnamefont {B.}~\bibnamefont {Kim}}, \bibinfo {author} {\bibfnamefont {E.}~\bibnamefont {Wallace}},\ and\ \bibinfo {author} {\bibfnamefont {C.}~\bibnamefont {Daraio}},\ }\bibfield  {title} {\bibinfo {title} {Modulation instability and wavenumber bandgap breathers in a time layered phononic lattice},\ }\href@noop {} {\bibfield  {journal} {\bibinfo  {journal} {Physical Review Research}\ }\textbf {\bibinfo {volume} {6}},\ \bibinfo {pages} {023045} (\bibinfo {year} {2024})}\BibitemShut {NoStop}%
\bibitem [{\citenamefont {Floquet}(1883)}]{GFloquet_1883}%
  \BibitemOpen
  \bibfield  {author} {\bibinfo {author} {\bibfnamefont {G.}~\bibnamefont {Floquet}},\ }\bibfield  {title} {\bibinfo {title} {Sur les {\'e}quations diff{\'e}rentielles lin{\'e}aires {\`a} coefficients p{\'e}riodiques},\ }in\ \href@noop {} {\emph {\bibinfo {booktitle} {Annales scientifiques de l'{\'E}cole normale sup{\'e}rieure}}},\ Vol.~\bibinfo {volume} {12}\ (\bibinfo {year} {1883})\ p.\ \bibinfo {pages} {47–88}\BibitemShut {NoStop}%
\bibitem [{\citenamefont {Rayleigh~(Strutt)}(1887)}]{JRayleigh_1887}%
  \BibitemOpen
  \bibfield  {author} {\bibinfo {author} {\bibfnamefont {J.~W.}\ \bibnamefont {Rayleigh~(Strutt)}},\ }\bibfield  {title} {\bibinfo {title} {On the maintenance of vibrations by forces of double frequency, and on the propagation of waves through a medium endowed with a periodic structure},\ }\href@noop {} {\bibfield  {journal} {\bibinfo  {journal} {The London, Edinburgh, and Dublin Philosophical Magazine and Journal of Science}\ }\textbf {\bibinfo {volume} {24}},\ \bibinfo {pages} {145–159} (\bibinfo {year} {1887})}\BibitemShut {NoStop}%
\bibitem [{\citenamefont {Sinha}\ and\ \citenamefont {Wu}(1991)}]{SSinha_1991}%
  \BibitemOpen
  \bibfield  {author} {\bibinfo {author} {\bibfnamefont {S.~C.}\ \bibnamefont {Sinha}}\ and\ \bibinfo {author} {\bibfnamefont {D.-H.}\ \bibnamefont {Wu}},\ }\bibfield  {title} {\bibinfo {title} {An efficient computational scheme for the analysis of periodic systems},\ }\href@noop {} {\bibfield  {journal} {\bibinfo  {journal} {Journal of Sound and Vibration}\ }\textbf {\bibinfo {volume} {151}},\ \bibinfo {pages} {91–117} (\bibinfo {year} {1991})}\BibitemShut {NoStop}%
\bibitem [{\citenamefont {Insperger}\ and\ \citenamefont {St{\'e}p{\'a}n}(2002)}]{TInsperger_2002}%
  \BibitemOpen
  \bibfield  {author} {\bibinfo {author} {\bibfnamefont {T.}~\bibnamefont {Insperger}}\ and\ \bibinfo {author} {\bibfnamefont {G.}~\bibnamefont {St{\'e}p{\'a}n}},\ }\bibfield  {title} {\bibinfo {title} {Stability chart for the delayed mathieu equation},\ }\href@noop {} {\bibfield  {journal} {\bibinfo  {journal} {Proceedings of the Royal Society of London. Series A}\ }\textbf {\bibinfo {volume} {458}},\ \bibinfo {pages} {1989–1998} (\bibinfo {year} {2002})}\BibitemShut {NoStop}%
\bibitem [{\citenamefont {Gill}(1951)}]{SGill_1951}%
  \BibitemOpen
  \bibfield  {author} {\bibinfo {author} {\bibfnamefont {S.}~\bibnamefont {Gill}},\ }\bibfield  {title} {\bibinfo {title} {A process for the step-by-step integration of differential equations in an automatic digital computing machine},\ }\href@noop {} {\bibfield  {journal} {\bibinfo  {journal} {Cambridge Philosophical Society}\ }\textbf {\bibinfo {volume} {47}},\ \bibinfo {pages} {96–108} (\bibinfo {year} {1951})}\BibitemShut {NoStop}%
\bibitem [{\citenamefont {Carnahan}\ \emph {et~al.}(1969)\citenamefont {Carnahan}, \citenamefont {Luther}, \citenamefont {Wilkes} \emph {et~al.}}]{BCarnahan_1969}%
  \BibitemOpen
  \bibfield  {author} {\bibinfo {author} {\bibfnamefont {B.}~\bibnamefont {Carnahan}}, \bibinfo {author} {\bibfnamefont {H.~A.}\ \bibnamefont {Luther}}, \bibinfo {author} {\bibfnamefont {J.~O.}\ \bibnamefont {Wilkes}}, \emph {et~al.},\ }\href@noop {} {\emph {\bibinfo {title} {Applied numerical methods}}},\ Vol.~\bibinfo {volume} {2}\ (\bibinfo  {publisher} {Wiley New York},\ \bibinfo {year} {1969})\BibitemShut {NoStop}%
\bibitem [{\citenamefont {Insperger}\ and\ \citenamefont {St{\'e}p{\'a}n}(2003)}]{TInsperger_2003}%
  \BibitemOpen
  \bibfield  {author} {\bibinfo {author} {\bibfnamefont {T.}~\bibnamefont {Insperger}}\ and\ \bibinfo {author} {\bibfnamefont {G.}~\bibnamefont {St{\'e}p{\'a}n}},\ }\bibfield  {title} {\bibinfo {title} {Stability of the damped mathieu equation with time delay},\ }\href@noop {} {\bibfield  {journal} {\bibinfo  {journal} {Journal of Dynamic Systems, Measurement, and Control}\ }\textbf {\bibinfo {volume} {125}},\ \bibinfo {pages} {166–171} (\bibinfo {year} {2003})}\BibitemShut {NoStop}%
\end{thebibliography}%


\providecommand{\noopsort}[1]{}\providecommand{\singleletter}[1]{#1}%
%

\end{document}